\documentclass[12pt,preprint]{aastex}
\slugcomment{Accepted for publication in ApJS 11 April 2001}

\shorttitle{Integral Field NIR Spectroscopy of Seyferts}
\shortauthors{Sosa-Brito et al.}

\usepackage{psfig}

\begin{document}

\title{Integral Field Near-IR Spectroscopy of a Sample \\
of Seyfert and LINER Galaxies I: The Data\altaffilmark{1}}

\author{Rafael M. Sosa-Brito, Lowell E. Tacconi-Garman, and Matthew D. Lehnert}
\affil{Max-Planck-Institut f\"ur extraterrestrische Physik, Postfach~1312,
85741~Garching~bei~M\"unchen, Germany}
\email{rmsb, lowell, mlehnert@mpe.mpg.de}

\and

\author{Jack F. Gallimore}
\affil{Department of Physics, Bucknell University, Lewisburg, PA 17837}
\email{jgallimo@bucknell.edu}

\altaffiltext{1}{Based on observations obtained with the
Anglo-Australian Telescope, Siding Spring, Australia, the European
Southern Observatory, La Silla, William Herschel Telescope operated on
the island of La Palma by the Isaac Newton Group in the Spanish
Observatorio del Roque de los Muchachos of the Instituto de Astrof\'{\i}sica
de Canarias, and the Calar Alto 3.5m which is operated by the
Max-Planck-Institut f\"ur Astronomie, Heidelberg, Germany, jointly with
the Spanish National Commission for Astronomy.}

\begin{abstract}
We present near-IR integral field spectroscopy of a sample of 31~Seyfert and
LINER galaxies which were selected both to span a wide range of nuclear
magnitudes and to possess roughly equal numbers of Seyfert type~1 and 2 nuclei.
Moderate resolution (R$\sim$1000; R$\sim$2000 for 3 cases) integral field
K-band spectra were obtained for all 31 galaxies in our sample and for
18~galaxies (R$\sim$1000; R$\sim$2000 for 4 cases) H-band integral field
spectra were also obtained.  In each case, we present nuclear, larger aperture,
and difference spectra with corresponding information about emission line
wavelengths, fluxes, and widths.  Line-free H- and K-band continuum images
as well as [Fe{\small II}]$\lambda$1.644$\mu$m, Br$\gamma$, and
\mbox{H$_2$~1-0~S(1)} emission lines are also presented.  In addition, we
provide extensive information about each galaxy obtained from the literature
that will be useful subsequently for characterizing the sample and for
comparison with our near-IR data.

\end{abstract}


\section{Introduction}

Active galactic nuclei (AGN) show an array of phenomenology
\citep[e.g.,][and references therein]{A93,K99}.  They are prodigious
producers of energy at all wavelengths and the reprocessing of their
intrinsic emission and competing sources of emission often complicates
the interpretation of their phenomenology.  What pieces of their
observed phenomenology is really intrinsic to the AGN and which are due
to extinction, scattering, reprocessing, or competing phenomenon like
the formation and evolution of massive stars?  To determine the
properties of the AGN implies that we must observe them over a wide range
of wavelengths and at high spatial resolution to be able to spectrally
and spatially segregate various emission mechanisms intrinsic to the AGN
and those which are extrinsic, and thus competing, sources of emission.
%Without such data, we will forever remain bewildered by the number of
%observed phenomena in nuclei of galaxies and make no progress in the
%theoretical interpretation of the emission characteristics of AGN and
%understanding and characterizing the types of environments which are
%able to create and sustain active galactic nuclei and supermassive black
%holes.

Recent work has demonstrated that star-formation near the nucleus can contribute
significantly to the optical--UV spectra of Seyfert nuclei.
%It has been recently realized that much of the phenomenology that
%historically had been attributed to the intrinsic properties of the AGN
%are actually due to star-formation on small spatial scales near the
%nucleus.
IUE spectra and HST imaging have been used to infer that only a
fraction of the UV continuum observed in Seyfert~2 galaxies is actually
due to a hidden AGN \citep[e.g.,][]{HKMCKKLR95, G-DHLMKWKK98,
HG-DLMKKKW97}.  In red optical spectra of Seyfert~2 galaxies,
\citet{TDT90} found that the equivalent widths of the Calcium triplet
lines ($\lambda\lambda$8498, 8552, 8662) were similar to those in the
bulges of normal galaxies.  In a moderately large aperture near-IR
absorption line study of a sample of Seyfert~1s and 2s, \citet{OOMM99,
OOKM95} have found that many Seyfert~2 galaxies show evidence for young
or intermediate age populations while such populations are rare (or
non-existent) in Seyfert~1 nuclei.  But overall they find that both
these young and old populations contribute significantly to the central
near-IR emission from Seyfert galaxies.

Near-IR spectroscopy and imaging have a unique role to play in advancing
our understanding of the AGN phenomenon.  This role reveals itself in
several important and unique aspects of near-IR data.  First, the
near-IR possess many emission lines and stellar absorption features that
are useful for determining the age and metallicity of the nuclear
stellar population \citep[like the strong CO bandheads in the H- and
K-bands; e.g., ][]{Sch98, OMO93}.  Second, extinction by dust declines
rather substantially from the UV through the optical out to the
near-IR\@. Thus observations in the near-IR allow one to probe depths
unreachable
with data obtained at shorter wavelengths.  Third, compared to the
continuum generated by stars, confusing sources of emission due to the
reprocessed emission from the AGN are also less significant in the
near-IR\@.  The emission from giant and supergiant populations peaks at
about 1.6 $\mu$m and is thus easier for a moderately young
($>$\,10$^7$\,yr) population to dominate all other emission mechanisms in the
near-nuclear environment, including reprocessing of nuclear continuum by dust
\citep[e.g.][]{Wardetal87}, compared to the optical or UV wavelength
regimes.  In the UV and blue optical, the dust scattering efficiency
is high enough and the recombination continuum
strong enough to easily dominate the emission from the underlying
stellar population.  Thus even with high spatial resolution and/or heavy
obscuration diminishing the light from AGN might not be enough to allow
one to investigate the stellar population of the nucleus unhindered by
contamination from AGN emission at these wavelengths.

Taking advantage of these unique characteristics of near IR data with
regards to investigating the stellar populations of active galactic
nuclei, we undertook a comprehensive near-IR (H- and K-band)
investigation of a large sample of Seyfert and LINER galaxies.
This paper presents the resulting  H- and K-band integral field
spectroscopic data from our sample of
31~Seyfert and LINER galaxies.  In addition to the strength of the near-IR data
generally, integral field spectroscopy uniquely allows us to investigate
the spatially resolved stellar populations of the nuclei of modest power
AGN and to investigate the distribution of the near-nuclear emission
line gas.  Thus unlike most studies conducted to date on the near-IR
nuclear properties of Seyfert galaxies, we do not have to exclusively
rely on large entrance apertures to investigate these properties.
The sample selection and characteristics is discussed in \S 2, the data
reduction in \S 3, some of the results are briefly discussed in \S 4
and what is known in the literature about each
of our sample galaxies is described in \S 5.
We defer discussion of the emission line and stellar
absorption line properties for now and limit ourselves to a description
of the sample and how the data were obtained and reduced.

\section{Sample Selection and Sample Characteristics}

\subsection{Sample Selection}

The galaxies chosen for this program were selected to span as broad a
range of nuclear magnitudes and optical line characteristics as possible
and still be classified as Seyfert or LINER galaxies.  These galaxies
were selected from the compilation of \citet{activecat}.  However, due
to limitations of the available observing time, the weather during the
scheduled observing time, and the locations of the various observatories
where the data were taken, the exact sample was a randomly selected
subsample of our entire list of active galaxies.
Our observed sample of galaxies is given in Table~1.

\subsection{Spectral Classification of the Nuclear Emission}

The initial classification criterion adopted to select our galaxies was
to adopt the classical spectroscopic definition of a Seyfert nucleus, namely, 
a ratio
[O{\small III}]$\lambda$5007/H$\beta$$>$3.  Overall, 22 of our observed
sample fits this criterion and in fact, 16 galaxies have [O{\small
III}]/H$\beta$ $>$ 5 (see Table~2).  However, six galaxies had an
observed ratio lower than 3, and a further two did not have reliable
measurements in the literature.  Although these sources do not meet our
strict selection criteria they are still useful for comparison purposes.
Specifically, these six are NGC\,1672, NGC\,1433, I\,Zw\,1, Mrk\,1044,
Mrk\,315, and NGC\,7582, while NGC\,4945 and Mrk\,231 did not have
reliable [O{\small III}]/H$\beta$ mesurements.  These cases
are discussed in more detail below.

After our initial selection, a more complete
investigation of the nuclear spectral classification was conducted by searching the
literature for line strength measurements.  To make the classifications
as robust as possible we searched for a wide variety of optical emission
lines but principally focused on [O{\small II}]$\lambda$3727, H$\beta$,
[O{\small III}]$\lambda$5007, [O{\small I}]$\lambda$6300, H$\alpha$,
[N{\small II}]$\lambda$ 6583, and [S{\small II}]$\lambda\lambda$6716,
6731.  Our specific selection of the lines to search for was based on
the criteria outlined by \cite{BPT81} and \cite{VO87}.  To make the
comparison of the emission lines and line ratios as fair as possible, we
tried where possible to ensure that the same apertures and slit position
were used for all lines and checked for uniform reddening correction of
the emission line ratios.  Further, they were checked to be approximately
corrected for the effects of underlying absorption in the line
strengths.  Finally, we sought uniform measurements of the emission line
widths to distinguish between Seyfert~1s and 2s.  While these were our
goals, due to a lack of data some compromises had to be made.  In most
cases, these measurements are internally consistent but not uniform
across the entire sample (e.g., the apertures are not the same from
galaxy to galaxy).

To classify our galaxies, our compilation of optical emission line ratios was used
together with diagnostic criteria from \cite{BPT81} and
\cite{VO87}.  For almost all of the galaxies in our sample the available data
allows us to determine their position in at least two of the diagnostic
diagrams.  In those cases where the
resulting classification differs from one diagram to the other,
greater weight was given to the classification derived from the [O{\small
III}]$\lambda$5007/H$\beta$ {\em vs.}~[N{\small II}]$\lambda$6583/H$\alpha$
(Figure~1)
and [O{\small III}]$\lambda$5007/H$\beta$ {\em vs.}~[S{\small
II}]$\lambda$6716\,+\,$\lambda$6731/H$\alpha$ diagrams, as these lines are
strong, and the ratios are only marginally affected by reddening.  The
diagrams involving [O{\small II}]$\lambda$3727/[O{\small
III}]$\lambda$5007 and [O{\small I}]$\lambda$6300 were given a lower
weight owing to the strong reddening uncertainties in the former case,
and the relative weakness of the line in the latter case.

In Table~2 we show the values for the ratios (corrected, with few exceptions,
for reddening
and underlying stellar absorption), over a range of apertures typically
of the order of a few arcseconds centered on the nucleus.  The Seyfert type is
chosen as 1, 1.5
or 2 according to the width of the hydrogen lines.  We opted not to further
subclassify our sample.
A classification of Seyfert~1.5 was chosen when clear signs of
underlying broad component on a narrow line was found on the literature.
In cases of doubt the conservative route was taken and the galaxy was designated
as either a ``pure'' Seyfert~1 or a ``pure'' Seyfert~2.  Widths broader
than about 1200~km~s$^{-1}$ were chosen for the classification as
type~1, less than about 800~km~s$^{-1}$ for type 2.  Between those two
values we used the narrowest and broadest measured line width to ``tip
the balance'' to one class or the other.  Our final classifications are
provided in Column~8 of Table~2.

There are a few sources which did not meet our original
[O{\small III}]$\lambda$5007/H$\beta$\,$>$\,3 selection criteria.
In the cases of NGC\,1433 and NGC\,1672, our adopted classification is LINER
\citep[see][and \S\S5.11 and \S\S5.13, respectively]{H80}.
I\,Zw\,1 and Mrk\,1044 have properties and line ratios like
``Narrow-Line Seyfert~1s''
(NLS1) and we adopt this classification (see \S\S 5.2 and \S\S 5.4, respectively,
for details)\footnote{It has been suggested \citep{HFSP97} that NGC\,4051
likely belongs to the NLS1 class, but we prefer the classification of a
classical Seyfert~1.}.
Mrk\,315 and NGC\,7582, on the other hand, show strong signs
of harboring Seyfert nuclei but probably also host significant
starbursts which greatly affect their nuclear emission line properties
(see \S\S 5.30 and \S\S 5.31).  For those cases we have adopted the classifications of
Seyfert~1.5/H{\small II} and Seyfert~2/H{\small II}, respectively.
The classification of NGC\,4945 is
particularly controversial.  A vigorous debate is underway about whether
or not NGC\,4945 harbors an AGN\@. For Mrk\,231 no line ratios are listed, anyway,
the Seyfert~1 classification was adopted based on the information available in the literature.
These arguments are discussed in
more detail in \S5.

Overall, our sample contains approximately equal numbers of Seyfert~1s
and 2s, with a few LINERs or possible LINERs.
Thus our final sample meets our competing goals of comparing
the near-infrared characteristics of active galaxies over a broad range
of nuclear magnitudes, optical emission line properties, and X-ray
properties (which means some ``weak AGNs'' like
LINERs should be included) and providing a robust comparison between
Seyfert~1s and 2s
(i.e., roughly equal numbers should be observed).

\subsection{Adopted Distances}

Since we are interested in comparing the physical properties of these
galactic nuclei in detail, the distances adopted in this paper require
discussion.  For the most distant galaxies, a simple Hubble
distance (D(Mpc) = V$_{3K}$/H$_0$) after correcting the observed
recession velocities to the microwave background restframe (V$_{3K}$) was
adopted.
A value for $\mathrm{H}_0$ of 75~km~sec$^{-1}$~Mpc$^{-1}$ was adopted for
this work, and the observed heliocentric velocities
(taken from the RC3 \citep{RC3}) have been corrected
following the work of \citet{lineweaver96}.  From COBE
observations, \citet{lineweaver96} find the velocity of the background
dipole to be $369\pm 2.5~km~sec^{-1}$ towards $\alpha = 11^h 11^m
57^s\pm 23^s$, $\delta = -7\fdg{22} \pm 0\fdg08$ (J2000).  Thus,
\begin{equation}
V_{3K}=V_{obs} + 76.2\,\cos\delta\sin\alpha -
358.1\,\cos\delta\cos\alpha - 46.4\,\sin\delta\,,
\end{equation}
where $V_{3K}$ is
the velocity relative to the microwave background, $V_{obs}$ is the
observed heliocentric velocity, and ($\alpha$,$\delta$) are the J2000
coordinates of the source.  The velocities in the microwave background
frame are listed in Table~1.

Of course, correcting for the solar peculiar velocity is only
appropiate for sources that are in the Hubble flow.  For example, for
galaxies with peculiar velocities as large as that derived for the
correction between heliocentric and microwave background frames would
imply an error of $\pm$5\,Mpc.  Since such an uncertainty would be
significant for galaxies within about 20\,Mpc, we have searched for
independent distance estimates for galaxies within 20\,Mpc with which to
compare with our simple Hubble law distances.
Our
adopted distances and the sources for those distances are presented in Table~1.
The mean adopted distances of our sample (see \S\S2.2) are $84\pm 26$\,Mpc,
$47\pm 19$\,Mpc, and $53\pm 17$\,Mpc for Seyfert classes 1, 1.5, and 2,
respectively.  Taken as a combined class, Seyfert~1 and 1.5 galaxies in our
sample have a mean distance of $68\pm 17$\,Mpc, in fair agreement with that
of the Seyfert~2 galaxies.

\subsection{Morphological Classification}

As one of the goals of this program is to study the nuclear regions of
Seyfert galaxies, we would therefore like to investigate how the
properties might depend on galactic morphological type.  The
morphological classifications are taken from RC3 \citep{RC3}.  Special
care should be taken when using the classifications of the more distant
galaxies since the photographic plates on which the RC3 classification
was based can be biased against structures on the furthest galaxies.
For completeness, the RC3 morphological types are shown in Table~1.

\section{Observations and Data Reduction}

\subsection{3D and ROGUE}

Imaging spectroscopy of the sample of Seyfert galaxies was obtained
using the MPE NIR imaging spectrometer 3D.\ \ Details of 3D are given by
\citet{Wetal96}, and only a brief summary is presented here.  3D uses a
mirror image slicer to divide a square field-of-view into 16~slitlets,
each 1~pixel wide and (roughly) 16~pixels long (see \S\ref{Data Reduction}),
which are subsequently
arranged end-to-end in a staircase pattern to produce a long
pseudo-slit.  This has the effect of reducing the two spatial dimensions
down to a single dimension.  This pseudo-slit is then spectrally
dispersed by passing it through a grism.  The final, two-dimensional
projection of the three dimensional (two spatial, one spectral) data is
imaged onto a $256\times 256$~pixel NICMOS~III detector.

%Spectral resolution achievable with 3D depends on the grism used for the
%observations.
3D has a suite of 5~grisms, two of which can be used on
any given observing run.  The wavelength range and achieveble resolution
for these grisms are listed in Table~3.  Prior to 1995 the data were
sampled at somewhat less than the Nyquist frequency, resulting in
spectral resolutions $\sim25\%$ lower than those listed in Table~3.
After that date, a piezo-driven mirror was implemented which is used for
half-pixel spectral dithering to realize the full resolution provided by
the grisms.

Starting in July 1994, 3D was used in conjunction with a tip-tilt
corrector ROGUE \citep{thatteetal95}.  This allowed us to achieve
spatial resolutions which were considerably better than the longterm
seeing values.  In addition, ROGUE carries with it a pixel scale
changing feature which allows observers to take full advantage of
changing seeing conditions.  The pixel size for all observations
(including those prior to July 1994) ranged from 0.25$^{\prime\prime}$
to 0.5$^{\prime\prime}$ resulting in fields-of-view between
4$^{\prime\prime}$ and $8^{\prime\prime}$ and which depended on the
seeing conditions and on which telescope the observations were taken.

\subsection{Details of the Observations}

Data were collected from November 1993 until April 1999 on various
telescopes located in both the northern and southern hemispheres.
Most of the data presented here were taken using the Anglo-Australian
Telescope. Tables~4 and 5 summarize the circumstances of the
observations.  The spatial resolution listed in Table~5 is that
achieved through the use of ROGUE, and represents a lower limit to the
actual NIR seeing at the time of the observations.

In a typical observation, the telescope was pointed alternatively to the
object and the sky in the pattern sky-object-object-sky.  Within this
pattern, half-pixel spectral dithering is performed between the two
object exposures.  As each on source exposure receives all the flux, the
total integration times listed in Table~5 are the sum of all of the
individual on source exposure times.  The time per individual exposure
was governed by our ability to successfully subtract the sky emission
from the source exposures.  Thus, H-band exposures were typically
60~seconds long, while K-band times ranged between 100 and 140~seconds,
depending on atmospheric conditions.

Since both the H- and K-bands are strongly contaminated by telluric
absorption, observations of stars with featureless, or weak-lined,
spectra in the H- and K-bands were observed to calibrate away the
atmospheric lines.  Typically, for H-band measurements
early type dwarfs or early F-type dwarfs were used, while for K-band
observations
late-F or early-G dwarfs were used.  These calibrators were observed
every 0.5--1~hour, depending on sky stability and the zenith distance of
the observations.

Most of our targets were observed for long periods during one or more
nights and all of these data were reduced independently for each night
of observation.

Flatfield frames were taken usually at the beginning and/or at the end
of the night.  Dome flats and internal lamp flats were taken, the first
consisting of exposures of an illuminated portion of the dome, the
latter consisted of exposures of a incandescent tungsten filament whose
spectral shape was easily removed from the data.  In practice, the dome
flats often had a spectral emissivity that was flat enough to provide a
reasonable flatfield and were often used exclusively to flatfield the
data.  Otherwise, combinations of internal and dome flats were used to
flatfield the data.  Exposures of an argon and neon arc lamps were made to
provide wavelength calibration of the H- and K-band respectively.  These
wavelength calibration exposures were taken every night and the final
wavelength calibration were checked against the wavelengths of strong
obvious night sky lines.  Dark exposures were obtained usually at the
end of each run, in order to remove the (small) dark current
contribution for each data frame.

\subsection{Data Reduction}
\label{Data Reduction}
The data were reduced using the GIPSY package \citep{gipsyref}, with
additional 3D-specific routines which were written in-house.  The data
were first corrected for (small) deviations from linear response before
subtracting the sky exposures from their corresponding object frames.

The sky-subtracted source frames are flatfielded, and dead pixel
corrected.  At that point the individual spectrally dithered pairs are
interleaved and the resulting frame is wavelength calibrated, using
calibration files generated from the arc lamp exposures.  From these two
dimensional representations of three dimensional information data cubes
are then generated.  The non-integer slitlet lengths require spatial
re-alignment by fractions of a pixel to obtain useful data cubes.
Sub-pixel interpolation is done using Fourier interpolation for
spatially well sampled data or by quadratic interpolation for less well
sampled data.

The data cube is then resampled onto a finer grid scale for more precise
centering of individual source frames with respect to each other.  The
pixel representing the center of each frame is chosen to be the peak
pixel within a broadband image (i.e.~an image generated by collapsing
the cube).  The fine-tuning of the peak position is then achieved by
assuming that the flux distribution very near the peak is Gaussian in
nature and the shifting the image such that the peak of a fitted
Gaussian lies at its center.  The set of re-registered source frames are
then combined into a final data cube.

The final reduction step is to remove the telluric absorption from the
object spectrum.  The data for the atmospheric calibrator are reduced in
exactly way as for the object.  The extracted one-dimensional spectrum
of the atmospheric calibrator is divided into the object data cube.

Calibration was done by
using aperture magnitudes available from the literature (see Table~5) and
appropriately scaling our spectra extracted over the same projected
aperture.

\section{Results}

We defer a detailed analysis of the full set of data to later papers.
Here we would like just to characterize the quality and gross features
of the data set.

Having three dimensional data for a large number of Seyfert galaxies
allows us access to a variety of diagnostic display methods.  The first
of these methods is one in which aggregate spectra over some
portion of the source are generated .  For example, a seeing-weighted
combination of individual spectra within the cube centered on the H- or
K-band continuum peak can be made.  Figure~2, shows 3~spectra for each galaxy:
one seeing-weighted, one which is the sum
of the data within an aperture of diameter 3\arcsec\ or
4\arcsec\ centered on the H- or K-band peak continuum emission and
another one which is the difference between the larger aperture and the
seeing-weighted spectra.  In each case,
the apertures chosen were determined by the seeing during the
observations and/or the field-of-view of the data.  For those cases where the field
is dominated by a strong central pointlike source, the seeing weighted
spectrum can be regarded as the ``nuclear spectrum.'' Measured fluxes of strong
H- and K-band emission lines, as well as broadband fluxes, derived from the
nuclear spectra, spectra of data within a 2\arcsec\ diameter aperture (not
shown), and the larger aperture (either 3\arcsec\ or 4\arcsec) spectra
are provided in Tables~6, 7, and 8.

Surprisingly, about half of the nuclear spectra are in fact
dominated by the stellar population of the nucleus, and not by the AGN
continuum.  The fraction of galaxies whose nuclear spectra are dominated
by a ``diluting continuum'' (relative to the stellar absorption line
strengths) is a function of Seyfert type, with Seyfert~1s preferentially
showing more dilution.
However, both Seyfert~1s and 2s
exhibit highly diluted nuclear spectra (e.g.~NGC\,1068, a Seyfert~2)
and both classes also exhibit
nuclear spectra that are dominated by their underlying bulge
populations (e.g.~NGC\,1566, a Seyfert~1).
The equivalent widths of the emission lines in most of the
nuclei are very low (Br$\gamma$ equivalent widths are rarely above
10\AA).  However, those nuclei with high Br$\gamma$ equivalent widths
are preferentially Seyfert~1s.  Many of the galaxies have blue
continua in both the nuclear and large aperture spectra suggesting that
they are dominated by stellar continuum emission, which must not be
heavily extincted.

The second means of looking at the data involves collapsing it
spectrally, either over the entire band (either with or without the
inclusion of lines) or over a line (with or without the inclusion of
local continuum).  In Figure~3 we present the spatially resolved maps
of the H- and K-band continua, and emission line maps of
[Fe{\small II}]$\lambda$1.644$\mu$m,
H$_2$ (including all detected H$_2$ lines in the K-band), and
Br$\gamma$.  The continuum maps were constructed by excluding
strong emission lines ([Fe{\small II}] in the
H-band, Br$\gamma$ and H$_2$ lines in the K-band).  In addition, the
K-band images include only wavelengths shortward of the CO bandheads at
2.29$\,\mu$m.  Thus, these are
``continuum only'' images.  The [Fe{\small II}]$\lambda$1.644$\mu$m,
H$_2$, and Br$\gamma$
emission line maps were constructed by including only those spectral
channels that included significant line emission.  The line maps have
been continuum-subtracted.  In addition, the line maps have been
spatially truncated such that only those pixels
that are $>3\,\sigma$ above zero are included in our plots.  Each map shows data for which
we are confident that the displayed morphology is statistically
significant.
Thus images not displayed for [Fe{\small II}] (when H-band data were
available), Br$\gamma$, or H$_2$ for any galaxy implies that the source lacked
significant emission in those lines or that the signal to noise ratio was not
good enough to create an image.
The H- and K-band continuum maps display the entire
spatial dimensions of the data cube and have not been truncated in size.

There is a wide range of morphologies in emission line maps, from point
like sources to more extended diffuse sources.  Interestingly, in the
continuum maps there are also a wide range of morphologies in the sense
that some images are obviously dominated by emission from a point
source, while other tend to be more diffuse and have no evidence for
harboring a point source.  A comparison of an aperture spectra and
continuum maps reveals that those sources with evidence for strong point
source emission also show diluted stellar continuum.
The presence of a nuclear point source is a function of
the Seyfert type, but some Seyfert~1's showing no such point source
and some Seyfert~2's exhibiting strong evidence for a point source.  This is
similar to what was found when considering the degree to which the nuclear
spectra were diluted. The distance distribution of Seyfert 2s is similar to
that of Seyfert 1s and 1.5s (see \S\S2.3), thus this result is independent of
any distance bias of the object types.

\section{Notes on Individual Objects}

In this section we provide a brief synopsis of the observational
characteristics of each object within our study as gleaned from the
literature, focusing particularly on the circum-nuclear properties of
each galaxy.  Special emphasis was given to other near-IR observations and
other direct evidence for the existence and properties of the active
nucleus.

\subsection{Mrk\,348 (=NGC\,262 = UGC\,499)}

This galaxy hosts a Seyfert~2 nucleus with strong emission lines
and underlying featureless continuum in the optical \citep{koski78}.
The host galaxy is an early-type spiral galaxy which is oriented nearly
face-on and has a nearby companion galaxy, NGC\,266 \citep{HSBS82, SSGH87}.
A broad H$\alpha$
component has been found in polarized light with a FWHM of 8400\,km~s$^{-1}$
\citep{MG90}.  \citet{VGH97} do not find broad Br$\gamma$ in
their IR spectrum but find Br$\gamma$ profile broader than that of
\mbox{H$_2$~1-0~S(1)}.  Our K-band spectrum superficially agrees
with theirs although the lower S/N of our spectrum does not permit more
conclusive comparisons.  A hard X-ray detection \citep{AKIH91} gives
N$_\mathrm{H} = 1.1 \pm 0.2 \times 10^{23}$\,cm$^{-2}$ and supports the idea
that Mrk\,348 harbors an obscured Seyfert~1 nucleus.  \citet{NB83} report
that its nuclear radio source consists of a compact core plus two knots
aligned along position angle 168$^\circ$, with a total size of about
0\farcs15.  They also report variation at 6 and 21\,cm on timescales of
months.  A high resolution HST image published by \citet{CAMSB96} shows
a linear structure of the emission in narrow-band [O{\small III}]$\lambda$5007
emission extended 0\farcs45 at a position angle of $\sim$155$^\circ$.

\subsection{I\,Zw\,1}

I\,Zw\,1 is sometimes classified as ``narrow-lined Seyfert~1'' (NLS1)
galaxy, due to its exhibiting a strong narrow component in some of its broad
emission lines \citep{OP85}.  In addition, its high optical luminosity
(M$_\mathrm{V}=-23.8$) is often used to claim that I\,Zw\,1 is the ``nearest
QSO.''  It has high X-ray luminosity \citep{KCU90} and strong optical
Fe{\small II} emission lines \citep{HO87} which are also typical of NLS1s.
\citet{SET98} find a strong presently decaying starburst located in a
circum-nuclear ring (1\farcs5 diameter) which contributes a large fraction
of the total (near-)nuclear light.  The optical emission line ratios (see
Table~2)
also indicate the presence of significant star-formation and its
composite starburst/AGN nature (we adopt the classification of
NLS1/H{\small II}).
\citet{HC90}
find that its host galaxy is a nearly face-on spiral
with a fainter companion to the west.

\subsection{Mrk\,573 (=UGC\,1214)}

Mrk\,573 is a Seyfert~2 galaxy showing strong high-ionization emission
lines \citep{koski78, TW92}.
No hard X-ray
flux was detected by {\em Ginga\/}
for this galaxy \citep{SD96}.  \citet{VGH97} find
that the nuclear near-IR emission from Mrk\,573 is dominated by stars
with faint Br$\gamma$ and \mbox{H$_2$~1-0~S(1)} emission.  Our
spectrum shows similar characteristics with a strong stellar
contribution and lack of significant diluting continuum.  The bar within
a bar, shown, for example in the K-band image of \citet{a-hsww98} (change from
PA $\approx$ 0$^\circ$
-10$^\circ$ in the outer to PA $\approx$ 80$^\circ$-90$^\circ$ in the inner isophotes
at 4\farcs5 of the center)
is also seen in our K-band image but our field-of-view is too small to
see the larger-scale bar.  \citet{TW92} find a biconical
radiation field roughly aligned with radio emission axis of Mrk\,573 found
by \citet{UW84}.  \citet{PDeR95} present high quality line maps, which
in [O{\small III}], for example, show clearly a biconical shape. They also find
two pairs of arclike emission line features that enclose the nucleus. The
innermost
of them are within 2\arcsec\ of the nucleus, which they interpret as bow shocks.
Unfortunately, our line maps are of insufficient signal-to-noise to
reveal any possible biconical emission in Br$\gamma$ or
\mbox{H$_2$~1-0~S(1)}.

\subsection{Mrk\,1044}

Mrk\,1044 is a NLS1 \citep[see][]{OP85} and it shows some evidence
for weak variability in its X-ray emission \citep{BBF96}.  It has strong
Balmer lines and the ratio I[O{\small III}]/I(H$\beta$) (1.67) is
uncharacteristically low for the narrow line region of a Seyfert galaxy.
However, such a low ratio is typical of objects classified as NLS1
\citep{OP85}.
\citet{RMS97} observe a narrow ($\mathrm{FWHM} = 2410$\,km~s$^{-1}$)
and a broad ($\mathrm{FWHM} = 10900$\,km~s$^{-1}$) component in
Ly$\alpha$, the narrow linewidth being compatible with our Br$\gamma$
measurement.
\citet{RS83}
obtained optical spectra and B-band imagery of Mrk\,1044.  They observe a
compact object in their B image, Fe{\small II} optical emission lines as well as
Balmer emission lines with faint broad component of
$\mathrm{FWHM} = 3400$\,km~s$^{-1}$ and with a narrower component superimposed.
A dominant
unresolved nucleus is observed by \citet{NMSG96} with the HST in the
near-IR, as is also the case in our H- and K-band images.

\subsection{NGC\,1068}

NGC\,1068 is the nearest Seyfert~2 galaxy and has been very
extensively studied.
NGC\,1068 has a polarized optical spectrum
similar to that directly observed
in Seyfert~1 nuclei, with $\mathrm{FWZI}\approx 7500$\,km~s$^{-1}$
for
the Balmer lines \citep{AM85}.
A compact stellar cluster has been found by
\citet{tqgmt97} and they show that its contribution to the total
bolometric nuclear (within a radius 2\farcs5)
luminosity is at least 7\%.  They also observe that 94\% of the light in
the K-band in the central 1\arcsec\ originates from a compact source which
they interpreted as hot dust emission.
The [O{\small III}] line emission morphology has a conical shape which extends
7\farcs5 at PA 35$^\circ$ with opening angle 45$^\circ$ \citep{evans91}.
A ring
of star formation (of outer diameter 36\arcsec) apparently powers half of the
mid-/far-IR luminosity while the other half comes from the Seyfert
nucleus \citep{PDeR93}.  Significant amounts of hot dense molecular gas
is shown by the \mbox{H$_2$~1-0~S(1)} line emission
extending over a region of 350 pc
around the nucleus \citep{rkcdgss91, blietz94} and thought to
be excited in the gas heated by UV radiation
or by X-ray photons from a central source \citep{rkcdgss91}.
Weak \mbox{H$_2$~1-0~S(1)} line emission, possibly due to shock excitation,
also
extends for 10\arcsec\ along the stellar bar \citep{DSW98,Scovilleetal1988}.
Our \mbox{H$_2$~1-0~S(1)} line map agrees very well with the bright inner
regions of the map from \citet{DSW98} (see their Figure~4).
The [Fe{\small II}]$\lambda$1.644$\mu$m forbidden
line has been mapped \citep{blietz94} and compared to the structure of
the radio elongated emission \citep{WU83}.  The
[Fe{\small II}]$\lambda$1.644$\mu$m and radio
emission line are co-linear, with the lobes of the radio
emission ``flaring out'' at the end of the significant
[Fe{\small II}]$\lambda$1.644$\mu$m
emission.  Forbidden [Si{\small VI}]$\lambda$1.962$\mu$m line has been
detected \citep{OM90} in this
object, however in our spectrum this spectral region is not covered.
\citet{EL88} present EXOSAT observations in the 2--10\,keV range. They detect
a source with a flat power law spectrum which resembles that of typical
Seyfert~1 galaxies. Within their observations they do not find evidence of
variations on timescales of 30~minutes to 4~years.
They use this as evidence that the direct view of the nucleus is totally
obscured and the X-ray flux observed
is seen only in scattered light.

\subsection{NGC\,1097 (=Arp\,77)}

NGC\,1097 was classified originally as
LINER on the basis of its optical spectra \citep{Keel83}, but
\citet{SWB96} re-classified it as a Seyfert~1 galaxy after observing the
appearance of broad Balmer line emission and a featureless blue
continuum.  Br$\gamma$ nuclear emission is not detected by \citet{KRLR00}
or in our spectrum of the nucleus, while H$_2$ emission is prominent
in their spectrum as well as ours.
In the K-band
image of \citet{KRLR00} a ring-like structure 10\arcsec\ from the nucleus
(outside our field-of-view) is
clearly seen.  This region corresponds to a ring of active
star-formation and appears to be located between two inner Lindblad
resonances \citep[see also][]{SWB96}.
\citet{PC96} present high resolution {\em ROSAT\/} X-ray
image of the nucleus and found that the circum-nuclear ring accounts
for 20\% of its total X-ray luminosity.
Hard X-ray emission from a point-like source
is reported by \citet{imftint96}
suggesting the existence of a low-luminosity AGN\@.
The nucleus of NGC\,1097 also
exhibits a complex system of jets discovered by \citet{WZ75} which are
15\arcsec\ wide and extend 5--19\arcmin\ from the nucleus.

\subsection{NGC\,1194 (=UGC\,2514)}

This Seyfert~1 galaxy has been little studied and thus not much is known
about it.  NGC\,1194 is a highly inclined spiral galaxy \citep{nilson73}.
A diffuse stream of gas towards the nearby companion galaxy UGC\,2517
\citep{nilson73} suggests a recent encounter.  It has warm IRAS
25--60 micron color
\citep{KL88, gkmgl92}.  Our K-band spectra is noisy and no emission
lines are detected, but CO absorption features are indeed present and
the continuum is red compared to that expected from a population of
giants or supergiants probably indicating either significant extinction
to the circum-nuclear stellar population or hot dust emission from the
active nucleus.  Our K-band image is consistent with NGC\,1194 having a
dominant central point-like component with an underlying, more extended
component.  These observations were, however, obtained under quite poor
seeing conditions and the quality of our data is low.

\subsection{NGC\,1275 (=Perseus\,A) }

NGC\,1275 is the central galaxy of the Perseus cluster with an optically
luminous nucleus which has been re-classified by \citet{veron78}
as a BL\,Lac object due to its weak line emission spectrum and to the
absence of broad lines as well as the variability and polarization
of its nucleus.
However \citet{HFSP97} report a broad H$\alpha$
component with FWHM 2750\,km~s$^{-1}$ and extremely wide wings (FWZI
19000\,km~s$^{-1}$) and we adopt the classification of Seyfert~1.5/LINER
(see Table~2).
\citet{RBMW81} observed that the active nucleus
is a hard X-ray source and it is variable on timescales of about one
year.  Intra-night micro-variability in the optical has been observed
\citep{PMM99}.  NGC\,1275 has a jet \citep{marr89, dhawan90, PBD83,
pedlar90} and counterjet \citep{VRB94} radio morphology.
NGC\,1275 is also cooling flow galaxy \citep[see, e.g.][]{HBBM89}.
This galaxy
has been studied in the near-IR by \citet{ksgtp00} who find that its
near-IR properties can best be described as a combination of dense
molecular gas, ionized emission line gas, and hot dust emission
concentrated on the nucleus (our spectra are red and show
significantly diluted stellar absorption features).
\citet{KS79} studied its two velocity systems and conclude that probably the
high-velocity emission line system (8200\,km~s$^{-1}$, several arcsec
northwest of the nucleus)
is excited by
hot stars while the low-velocity filamentary emission line structure
(5300\,km~s$^{-1}$, the systemic velocity of NGC\,1275)
can be explained by either shocks or photoionization from the Seyfert nucleus.
There is also a central population of young massive
star clusters \citep{holtzman92}, roughly 15 of which lie within our
field-of-view.

\subsection{NGC\,1365}

NGC\,1365 is a spiral barred galaxy located in the Fornax cluster.
It exhibits a conical [O{\small III}] emission morphology \citep{kjlb97}, and
shows
starburst activity \citep[e.g.,][]{TDW93} with star forming
regions concentrated in ``hot spots'' around the nucleus, outside of our
field-of-view.
\citet{kjlb97} observed this galaxy with the FOC onboard HST
and  detected numerous bright super star clusters
associated with the circum-nuclear star forming regions
\citep{SFN99}.  \citet{RMB00} observed NGC\,1365 with {\em BeppoSAX\/} and found
its 4--10\,keV emission to be highly variable over the course of the
observations (varying by a factor of 2 over 50,000 seconds).
Broad
hydrogen emission lines have been seen on the nucleus \citep[see for
example][]{SKSM99, EP82} indicating the presence of an AGN\@.  Our data
also shows a broad nuclear Br$\gamma$ line, together with a red continuum
with relatively weak stellar absorption features.  VLA observations
\citep{SJL95} reveal a radio jet from the nucleus in the direction of
the optical emission line cone.

\subsection{NGC\,1386}

The presence of the AGN in this galaxy is revealed by
both water maser emission and their associated velocities \citep{braatz97},
as well as
by a strong Fe-K emission line seen in {\em ASCA\/} observations
\citep{imfti97}.  An infrared excess is observed in the L- and N-bands
\citep{SHAB86} but not in the J-, H- and K-bands.  Our K-band
spectra shows little evidence for a continuum diluting the relatively strong
stellar absorption features.  Speckle H$\alpha$ observations
show an elongated structure
$\sim$3\arcsec\ long (centered on the nucleus) along a position
angle of $\sim$30$^{\circ}$ consisting of several knots \citep{mauder92}.
Our broadband H- and K-band maps do show
similar elongation to that found by \citet{mauder92}.
\citet{UW84b} found that the morphology of the nuclear radio continuum is
extended towards the southwest along galaxy major axis (PA 55$^\circ$)
by about 400\,pc.
Evidence for nuclear outflow along the same axis is found by \citet{WWB91}.

\subsection{NGC\,1433}

NGC\,1433 has been classified as Seyfert~2 \citep{VV86} but we prefer to
classify it as a LINER on the basis of its emission line ratios from the
literature (Table~2).  No direct sign that NGC\,1433 harbors an AGN has
been found, but it does possess a rich morphology with two bars with
position angles which differ by 64$^{\circ}$ \citep{buta86} and three
concentric rings, the most inner one of 18\arcsec\ diameter \citep{buta86}.
\citet{JCA97}
detected the inner bar in the near-IR which is seen in our images
with a similar morphology.  H{\small II} regions are observed along the inner
ring \citep{buta86}, but this region lies outside of our
field-of-view.  \citet{harnett87} found in their radio observations that
the radio emission peaks on the position of the optical nucleus and that
weaker radio emission extends along the large scale bar.

\subsection{NGC\,1566}

This galaxy possesses weak emission lines for a Seyfert \citep{OSW74}.
HST WF/PC1 observations of NGC\,1566 made by \citet{ketal90} show that the
emission line gas has a point source contribution with a component that
extends to $<$0\farcs7, while subsequent HST continuum imagery reveals
spiral dust lanes within 1\arcsec\ of the nucleus \citep{GHG97}.  Its optical
spectrum is that typical of a Seyfert~1 galaxy \citep{ketal90}.
However according to \citet{HP80} its nuclear spectra resembles more that of an
H{\small II}
region than that of a Seyfert galaxy except for the broad emission wings
at H$\alpha$.
A strong X-ray point source with weaker extended emission was found
through {\em ROSAT\/} observations \citep{EBHVPER96}.  Multi-wavelength nuclear
variability (from X-ray to near-IR) has been reported by \citet{BAGP92}
with X-ray emission variations up to 40\% within timescales less than a month.
A prominent large scale bar is seen in near-IR images \citep{HS83}.  Both large
scale and nuclear radio emission is seen in NGC\,1566 \citep{harnett84,
harnett87}.

\subsection{NGC\,1672}

We adopt for this galaxy, based on the line ratios from the literature,
a classification of LINER galaxy.
NGC\,1672 shows X-ray emission from the ends of its prominent bar, from two
other off-nuclear regions, and from
an extended nuclear source \citep{DBH99, NBHI00}.
The nuclear emission, which is the dominant soft X-ray source,
shows no variability \citep{NBHI00}.  \citet{NBHI00} note that the bulk of the
2--10\,keV and 5--10\,keV comes from the off-nuclear source X-3.  Further,
they do not
find evidence for an absorbed nuclear X-ray source and conclude that if
NGC\,1672 harbors a luminous Seyfert nucleus, it must be obscured even in
the X-rays (requiring column densities $>$ 2 $\times$ 10$^{24}$
cm$^{-2}$).  H$\alpha$ knots are found in a ring structure within 10\arcsec\ of
the nucleus, which is likely to be a region of recent star-formation
\citep{evans96}.  The velocity field around the nucleus indicates a 9
$\times$ 10$^8$ M$_{\sun}$ within the inner 1\farcs8 \citep{dcdg99} which is
suggestive of a heavily obscured AGN (but is certainly not conclusive).
\citet{MMSO94} detect \mbox{H$_2$~1-0~S(1)}, \mbox{H$_2$~1-0~S(3)}
but no [Si{\small VI}]$\lambda$1.962$\mu$m
in their K-band spectrum.  Our K-band spectrum shows no H$_2$, despite our
achieving a noise level more than a factor of three lower than that of \citet{MMSO94}.

\subsection{Mrk\,1095 (=UGC\,3271 = Ark\,120)}

The nucleus of Mrk\,1095 exhibits a typical Seyfert~1 spectrum \citep{RS91} whose optical
continuum and emission line show variability over times scales of years
\citep{PWBHPW98}.
A dominant unresolved nucleus is reported on HST observation by \citet{NMSG96}.
The 3D data shows very broad Br$\gamma$ emission centered on the nucleus and
a very red spectrum.  The latter can be interpreted as hot dust emission
diluting the stellar continuum.
This is in qualitative agreement with \citet{OOMM99}, who use
a technique of fitting stellar template spectra to restricted portions of the H-
and K-band spectra and report stellar contributions of 50\% and 25\% at
1.62\,$\mu$m and 2.29\,$\mu$m, respectively.

\subsection{NGC\,2110}

HST imaging spectroscopy of this source reveals both a narrow, 1\arcsec\ long
jet/region of [O{\small III}] emission extending to the north of the
nucleus and an S-shaped H$\alpha$ emission region within the inner
4\arcsec\ \citep{metal94}.  A similar S-shaped morphology is seen in our
[Fe{\small II}] map.
Radio emission extends symmetrically $\sim$2\arcsec\ north and
south of the nucleus \citep{UW83} while [O{\small III}] and
H$\alpha$+[N{\small II}] images show more extended emission which is also
elongated north-south
\citep{WBU85, Pogge89}.
Hard X-ray observation of its nucleus are consistent
with an obscured Seyfert~1 nucleus \citep{malaguti99}.
\citet{qetal99} imaged NGC\,2110 with NICMOS and find the
molecular hydrogen emission to be extended.
In agreement with these data are those of \citet{SWWW99}, who observed
NGC\,2110 with a long slit
in the near-IR and found extended [Fe{\small II}]$\lambda$1.257$\mu$m and
\mbox{H$_2$~1-0~S(1)} emission.
Our data lack sufficient
signal-to-noise to see extended \mbox{H$_2$~1-0~S(1)} emission.
Storchi-Bergmann et al.~conclude that an important source of
exitation for [Fe{\small II}]$\lambda$1.257$\mu$m may be shocks driven by the
radio jet and while the H$_2$ is excited by the central X-ray source.
They report that the dynamical center is
displaced with respect to the peak of near-IR continuum and
find signatures of hot dust emission (in agreement with our K-band
spectra).

\subsection{NGC\,3079}

NGC\,3079 is a nearly edge-on galaxy (our K-band image is also elongated at the
same position angle as the galactic disk) which has
been well studied.
NGC\,3079 has been
classified as a LINER \citep{H80}, but we prefer a Seyfert~2 classification
based on line ratios from the literature (Table~2).
\citet{LH96} and \citet{vcbtfs94} find evidence that the nucleus
is driving a galactic-scale outflow (a ``super-bubble'').
On the
nucleus, large concentration of CO gas has been found \citep{SS94} and
a central starburst appears sufficiently strong to produce the outflow
\citep{vcbtfs94},
though the latter authors cannot exclude the presence of an AGN\@.
A detection of broad H$\alpha$, consistent with the nuclear outflow, has been
reported by \citet{stauffer82}.
\citet{IWHA98} image the central region
of this galaxy in the near-IR (J, H, K and the \mbox{H$_2$~1-0~S(1)} line).
They
claim that its extremely red infrared colors can only be explained with
hot dust emission, and support that argument by comparing the depths
of the K-band CO absorption lines from \citet{HIGW95} with those from
\citet{AGC89} and \citet{LR-V92}.  \citet{IWHA98} infer based on this
comparison that up to 25--30\% of the emission within the central
$3\arcsec\times 3\arcsec$ can be attributed to hot dust.  However, our K-band
spectrum, at higher spectral resolution than that by \citet{HIGW95}, shows
CO depths 25--30\% deeper than theirs, and thus does not reveal a
hot dust signature.
\citet{HIGW95} conclude that the H$_2$ near-IR emission lines are shock
excitated in the outflow and that the outflow has an active nuclear origin.
\citet{PTV98} resolve the X-ray emission from the inner
$20\arcsec\times30\arcsec$ and find it to be coincident with
the optical super-bubble.  A point-source active  nucleus
may contribute to the X-ray emission.
In addition, they argue that an AGN interpretation is favored for the
high X-ray to optical luminosity ratio.

\subsection{NGC\,3227 (=Arp\,94)}

NGC\,3227 is classified as Seyfert~1.5 and is interacting
\citep[see e.g.][]{mundell95} with NGC\,3226, a dwarf elliptical companion.
\citet{MHPMKA95} obtained ground based [O{\small III}] imaging showing an
extended (7\arcsec\ to the NE of the nucleus, PA $\approx$ 30$^\circ$)
structure non-aligned with their radio map (6 and 18 cm) showing a
0\farcs4 double radio source at PA $\approx$ $-$10$^\circ$ (the difference in
position angles between the radio and optical structures is $\sim$40$^\circ$).
\citet{SK96} present an archival [O{\small III}] HST image made with a shorter
exposure time than that by \citet{MHPMKA95} showing a compact nucleus and
extended (0\farcs9) emission (PA 15$^\circ$).  \citet{AM94} present results of
2-dimensional optical spectroscopic observations of the inner kiloparsec of this
galaxy showing both AGN and H{\small II} region-like emission line properties.
NGC\,3227 shows strong spectral variability in the optical
\citep[see][]{setal94} and in X-ray \citep{gmtypnn98}.  A NICMOS image of
\citet{qetal99} shows resolved \mbox{H$_2$~1-0~S(1)} line emission which our
data also reveals.  The H$_2$ emission is elongated in the same
way as the compact $^{12}$CO~(1-0) millimeter interferometric map of
\citet{SET00}, and our H$_2$ peak and the CO peaks ($^{12}$CO~(1-0)
and $^{12}$CO~(2-1)) are offset from the nuclear position, though the H$_2$
offset is only $\sim$0\farcs3, smaller than that observed in the interferometric
CO observations.  Speckle images
obtained by \citet{WNM98} in the H- and K-band show an unresolved point source
at or near the diffraction limit of the Palomar 200-inch telescope on top of an
extended region of continuum emission (in agreement with our broadband images)
and also find that the colors of the nucleus are consistent with hot dust
emission, as also suggested by the spectra shown in Figure~2.

\subsection{NGC\,3393}

\citet{DPW88} studied the optical and UV spectra of the nuclear region of
NGC\,3393 and find no evidence for internal reddening.  Further, they find that
the spectral type of the dominant stellar population is sufficiently old
that it does not contribute to the UV continuum, but rather contributes
$\geq$90\% of the optical continuum.
Their analysis of the IRAS fluxes shows that it can be fit with a combination of
two relatively cool components, one at 130\,K and the other at 30\,K\@.
For comparison, our K-band spectrum shows no clear evidence for hot
dust emission \citep[see also][]{a-hsww98}.
\citet{ferguson97}, and later \citet{CBFNW2000}, show optical line strengths
and ratios which clearly distinguish this galaxy as a Seyfert~2.
The weak 20-100\,keV emission seen with {\em BeppoSAX\/} has been modelled by
\citet{MSBDCMRZ98} who favor a cold reflection dominated (Compton thick) model.

\subsection{NGC\,3783}

NGC\,3783 is a nearly face-on spiral galaxy hosting a very bright
\citep[e.g.][and references therein]{aetal95} Seyfert~1 nucleus (note the
very broad Br$\gamma$ emission in our K-band spectra).
\citet{aetal95} find that if the $\alpha=1$ power law that fits the 0.1\,MeV --
0.1\,keV emission is extended to longer wavelengths, the resulting IR excess
is explainable by a $\sim$60\,M$_\odot$ mass of dust with temperatures in the
range 200--1500\,K\@.
Our K-band spectrum also shows evidence for substantial hot dust emission.
\mbox{H$_2$~1-0~S(1)} line emission from the central $3\farcs 4\times 6\farcs 8$
has been marginally detected \citep{KNG89}, consistent with our upper limit.

\subsection{NGC\,4051}

NGC\,4051 has ben referred to as the least luminous ``classical'' Seyfert~1
\citep{HFSP97, weedman76}.
HST archive [O{\small III}] emission line imaging \citep{SK96}
reveals an unresolved nuclear
source with fainter extended (1\farcs2) emission along a position angle of
100$^\circ$, similar to the radio continuum structure observed by \citet{UW84}.
\citet{Veilleux91}, in his study of the
structure and kinematics of the Narrow Line Regions in Seyfert galaxies, notes
that the line profiles of {\em all of the observed forbidden lines\/} show
pronounced blue wings.  There are no corresponding red wings, which led
him to propose a model including outflow towards us plus dust obscuration
to extinct the flow away from us.
Hot dust has also been invoked to explain the mid- and far-infrared continuum
emission \citep{RPLM96,CV99}; the former authors differentiate between hot dust
heated by the AGN and somewhat cooler dust which is heated by star formation.
Variability of NGC\,4051 occurs on a number of different timescales.
\citet{setal93} report an outburst at 2.2\,$\mu$m of more than a factor of 2 in
6 months in 1992, over which time the UV emission appears to be less variable.
X-ray variability with timescales as short as a few hundred seconds has been
observed by {\em EXOSAT\/} \citep{Lawrenceetal85}, while \citet{detal90} report
on observations of strong X-ray variations within timescales of tens of
minutes.
Finally, \citet{singh99} present a high resolution X-ray map for the nuclear
region (5\arcsec--10\arcsec) of NGC\,4051 and interpret it as due to central
(nuclear) activity plus an extended starburst component.

\subsection{NGC\,4258 (=M\,106)}

NGC\,4258 contains a nuclear thin disk in Keplerian rotation, traced by emission
from water masers \citep{metal95}.  The implied enclosed mass density,
3.6 $\times$ 10$^7$ M$_\sun$ within 0.14\,pc of the nucleus \citep{Hetal96},
consistent with a supermassive central black hole.
In spite of the clear presence of a nuclear black hole, NGC\,4258 is a
relatively inactive system.  Both the nuclear H$\alpha$ and radio continuum
luminosities are low, and the nucleus was originally undetected in X-ray
observations made with the {\em Einstein Observatory\/}
\citep[][and references therein]{ford86}.
Subsequent observations with {\em ASCA\/} have
revealed a low luminosity obscured active nucleus \citep{mfiklmssty94}, and
\citet{fs85} have observed very weak broad H$\alpha$ emission.
Our K-band image (see Figure~3) shows an unresolved central source, and recently
\citet{CBENSMR00} have shown this source to be unresolved down to
0\farcs2 FWHM resolution.
These authors have also fit their NIR flux densities with a single, non-thermal,
powerlaw ($\nu^{-1.4\pm0.1}$), using a foreground screen extinction correction
corresponding to an A$_\mathrm{V}$ of 18~magnitudes.

\subsection{NGC\,4639}

This galaxy contains many H{\small II} regions which lie both at the
ends of a stellar bar and in a ring 30\arcsec\ from the
nucleus \citep{G-DPTVR-E97}.  The AGN itself appears to contribute very little
(3\%) to the total ionizing luminosity \citep{G-DPTVR-E97}.
Owing to their detection of broad hydrogen emission lines
(FWHM\,=\,3600\,km~s$^{-1}$ for H$\alpha$),
\citet{HFSP97} classified this galaxy as Seyfert~1 (although low-luminosity).
Further, they report an H$\alpha$
flux variation of $\sim$10\% between two observations a year apart, though caution
should be used in interpretting this due to the sensitivity of the result on
detailed stellar light removal.
The Seyfert nucleus, which is pointlike in the {\em ROSAT\/} soft X-ray band,
has a low X-ray luminosity but
its X-ray spectral properties are very similar to more powerful Seyfert
nuclei \citep{KDHFHR95}.
\citet{HPTKSYK99}, compared {\em ASCA\/} data with archival {\em Einstein\/}
and {\em ROSAT\/} data an find variations on scales of months to years, with
some suggestion of variability on timescales of $\sim$10$^4$ seconds from
the {\em ASCA\/} data alone.

\subsection{Mrk\,231}

This galaxy is one of the most luminous objects in the local universe \citep{sanders88} --
it is a member of the ultraluminous infrared galaxy class of objects, with
$L_{\mathrm IR}\sim 3.5\times 10^{12}\,L_\odot$ \citep{LCM94}.
The presence of an AGN is revealed by many observations.  In the radio,
\citet{prefos83}, \citet{NU88}, and \citet{UWC99} have reported on the
existence of a very compact, variable, self-absorbed source in the nucleus.
That, in combination with the observed parsec-scale jet
\citep{NU88,UWC99,TSUC99},
clearly signals the presence of an AGN\@.  Moreover, UV and optical
polarization observations \citep{goomil94,smithetal95} point towards the
existence of a central, highly polarized, nonthermal source.
Its X-ray emission is weak
compared to normal Seyfert~1 galaxies and although hard X-ray emission
is detected, no variation is observed \citep{nakagawaetal99,turner99}.
Our K-band imaging spectroscopy reveals the presence of an unresolved central
source.  \citet{WNM98} have used speckle techniques at 1.6 and
2.2$\,\mu$ to demonstrate that this nuclear source is unresolved down to
roughly the diffraction limit of the Palomar 200-inch telescope.
Mrk\,231 is also host to several blueshifted broad absorption line systems
seen in UV data \citep[see references in][]{turner99}, the shapes and
velocities of which are similar to those seen in BAL QSOs
\citep{rudyetal85}.
Strong (circumnuclear) starformation is also indicated by a number of
observations.
The VLBA and VLA observations of \citet{TSUC99} show, in addition to the
nuclear jet, an extended component of size 100--1000\,pc, which they
associate with a region of starformation.  They also report the presence of
4 yet-to-be-confirmed radio supernovae.
Regions of active circumnuclear star formation have been seen in the
optical observations of \citet{HN87} and \citet{HK87}.  Groundbased J-, H-,
and K-band observations with arcsecond seeing revealed the presence of a
second source 3.5\,arcseconds south of the nucleus \citep{ASSMGL94}.  This
source was originally described as a second nucleus, but higher resolution
{\em HST\/}/WFPC2 observations of \citet{suraceetal98} have shown that to
be a string of starforming knots.
Millimeter interferometric observations CO in Mrk\,231 reveal the presence
of a dense, rotating, compact, nuclear disk of molecular gas \citep{DS98},
which the authors argue fuels a circumnuclear starburst.  Further,
\citet{UWC99} note that the $L_{\mathrm X}/L_{\mathrm IR}$ ratio implied by
the {\em ASCA\/} observations \citep{nakagawaetal99} is more consistent
with that seen in starburst galaxies than for AGN\@.
\citet{KCTK97} present non-ROGUE-assisted K-band 3D integral field spectroscopy
and find extended circumnuclear emission from thermally excited hot
molecular gas as traced by the H$_2$ lines.  Although present, the
CO-bandhead stellar absorption feature at 2.29$\,\mu$m appears to be
sitting on the steep slope of a rising hot dust continuum.  The results
presented here, although of poorer signal-to-noise, are consistent with
these findings.

\subsection{NGC\,4945}

There is a debate in the literature whether NGC\,4945, one of the brightest
infrared galaxies in the sky ($L_{\mathrm 8-1000\,\mu m}=2.95\times
10^{10}$\,L$_\odot$, \citet{Spoon2000}, and references therein),
harbors an AGN\@.  The side taken is a strong function of wavelength observed.
Firm evidence for the presence of an AGN comes primarily from (hard) X-ray
observations.  \citet{IKAKMTON93} report the results of {\em Ginga\/}
observations in which the 2--10\,keV band has a powerlaw emission spectrum with
photon index $\sim$1.7, an absorption-corrected luminosity of $3\times
10^{42}$\,ergs~s$^{-1}$, and strong variability on a timescale of several
hours.  Similar results based on {\em BeppoSAX\/} observations are presented by
\citet{Guainazzietal2000}.  \citet{DMS96} use OSSE data to determine that
NGC\,4945 is the second brightest Seyfert at hard X-ray energies (the
brightest being NGC\,4151).  Those authors aptly note that the AGN in NGC\,4945
sits behind an obscurring column of $\sim 5\times 10^{24}$\,cm$^{-2}$
\citep{IKAKMTON93}, rendering it invisible at all energies below 10\,keV
\citep[see also][]{Spoon2000}.
Spectroscopy at mid-IR wavelengths with {\em ISO\/} \citep{Spoon2000}
suggest that at least 50\% of the bolometric luminosity of NGC\,4945 derives
from massive star-formation, which is heavily obscurred by 36$^{+18}_{-11}$
magnitudes of visual extinction.  \citet{Spoon2000} see no evidence from the
mid-IR
lines for excitation attributable to an AGN\@.
The starburst in this galaxy has produced a superwind which has evacuated a
cone-shaped structure, perpendicular to the plane of the galaxy
\citep{HAM90, LH96, MvdWKMO96}.
\citet{qetal99} and \citet{Marconietal2000} both present {\em HST\/} NICMOS
imaging spectroscopy of the 1-0~S(1) line of H$_2$ at 2.12$\,\mu$m.  The
H$_2$ emission is extended, and traces the inside edges of the superwind
bubble \citep{Marconietal2000}.  Our H$_2$ line map also shows extended
emission which agrees well with the bright central emission seen by
\citet{Marconietal2000}, though our field of view is insufficient to trace out
the full extent of the superwind cavity.
Our H$_2$ flux within a 4$^{\prime\prime}$
aperture is consistent
with $1.1\times 10^{-13}$\,ergs~cm$^{-2}$~s$^{-1}$ observed within a
$6^{\prime\prime}\times 6^{\prime\prime}$ nuclear aperture by
\citet{Marconietal2000}.  Our [Fe{\small II}] map shows extended
bubble-like
emission similar to, but smaller in scale than, that found
by \citet{Marconietal2000}.

\subsection{Mrk\,273 (=UGC\,8696)}

Mrk\,273 is an ultraluminous infrared galaxy (log L$_{IR}$ = 12.11
L$_{\sun}$).  That an AGN at least contributes to the excitation of the
nuclear region is indicated by the optical
spectroscopy of \citet{koski78,sanders88} (see also the K-band spectrum of
\citet{VSK99}), as well the strong high excitation
lines observed by {\em ISO\/} \citep{genzel98}.  The latter work, however,
also shows
Mrk\,273 to have a relatively strong 7.7\,$\mu$m PAH feature, suggesting
that of order 50\% of the nuclear gas excitation is due to a starburst.
The starburst nature of the nucleus is also demonstrated by the K-band spectrum
presented by \citet{GJDS95}, though this spectrum does extend beyond the red
nuclear region to the bluer disk \citep{Scovilleetal2000}.  In contrast,
our K-band spectra extracted over the central 4$^{\prime\prime}$ are much
redder with apparently weaker stellar features, and are consistent with the
spectrum of \citet{VSK99} over a similar region.
The {\em ASCA\/} observations of Mrk\,273 indicate that it contains a Seyfert
nucleus of at least moderate luminosity \citep{Iwasawa99}.
Observations of the nuclear region at higher spatial resolution than those
discussed above have recently
been made in the NIR\footnote{\citet{KLYRRDN97} fail to
detect the SE nuclear source seen in the {\em HST\/} NICMOS image at
0.22$^{\prime\prime}$ resolution \citep{Scovilleetal2000}.  This is likely due
to a relatively low Strehl ratio, resulting in little power being in the
diffraction-limited core of their point spread function.  The true resolution
of these data is likely not substantially better than their quoted seeing
value of 0.6$^{\prime\prime}$.  This would also account for the fact that in
their image the brighter nucleus is the northern one, rather than the
southwestern as seen
by \citet{Scovilleetal2000}.} \citep{KLYRRDN97, Scovilleetal2000} and in the
radio by \citet{KLYRRDN97} and \citet{CT00}.  These measurements collectively
show the ``nucleus'' to be composed of three separate sources, denoted N, SE,
SW.  Of the two bright NIR sources, N, and SW, only N is detected in the radio.
\citet{KLYRRDN97}, \citet{Scovilleetal2000}, and \citet{CT00} all discuss
the nature of the three sources, and come to somewhat conflicting viewpoints.
\citet{KLYRRDN97} argue in favor of the northern source being an AGN, while
\citet{Scovilleetal2000} and \citet{CT00} present compelling arguments for that
component to be starburst in nature.  Further, \citet{CT00} interpret their
H{\small I} absorption line data from the northern nucleus as indicative of
a gaseous disk in rotation, with the inferred enclosed mass
($2\times 10^9$\,M$_\odot$) being fully consistent with the molecular gas
observed by \citet{DS98}.  \citet{Scovilleetal2000} posit that the
AGN is the SW nucleus, which is very red and unresolved in their NICMOS image,
while \citet{CT00} do not strongly favor either the SE or SW component being an
AGN\@.

\subsection{Mrk\,463 (=UGC\,8850)}

Mrk\,463 consists of two nuclei, Mrk\,463W and E, separated by
about 4\arcsec\ \citep[see also our K-band image]{Adams77}.  Without
distinguishing between the two nuclei, \citet{LVG99} use newly developed
mid-infrared tools to classify Mrk\,463 as an AGN\@.
Indeed, Mrk\,463E is classified by most authors as a Seyfert~2
\citep[e.g.][]{SO81, HN89}, and shows in {\em HST\/} direct imaging
what was initially described as a 0.84$^{\prime\prime}$ long optical jet
directed towards the south of the nucleus \citep{UCFRAEC93}.
The Seyfert~2 nucleus appears as a Seyfert~1 in reflected polarized light
\citep{MG90}, and more recent {\em HST\/} imaging polarimetry has revealed
that the ``optical jet'' is in fact a cone of polarized light extending
{\em northward\/} from the Seyfert~1 nucleus \citep{TUATFK96}.
In addition, the eastern
component is seen to have broad Pa\,$\alpha$ emission \citep[][consistent
with our data which includes only the red side of the line]{VSK99}.
Mrk\,463W, on the
otherhand, is classified as either a Seyfert~2 \citep{SO81,MGSGNM91}
or as a LINER \citep{Blanco91}.  Although there are arguments in favor of
other classifications, we adopt the Seyfert~2 classification for both nuclei.
The unresolved nuclei seen in our
K-band data have been shown to be unresolved at higher spatial resolution
\citep{MGSGNM91}.

\subsection{Circinus}

Circinus lies close to the galactic plane, within a window of relatively
low (A$_V$ $\approx$ 1.5 mag.) interstellar extinction \citep{Fetal77}.
It displays
several characteristics of a typical Seyfert~2 nucleus: the observed optical
line ratio \citep{OSMM94}, an ionization
cone observed in [O{\small III}]$\lambda$5007 \citep{mmoo95} with the
corresponding counter-cone appearing in the 1.97$\,\mu$m [Si{\small VI}]
line \citep{MA-HAQRRT-G2000},
narrow prominent
coronal lines in its optical/near-IR spectrum \citep{OSMM94,Moorwoodetal96,
MKTG98}, and broad
(FWHM$\approx$ 3300 km s$^{-1}$)
H$\alpha$ emission detected in polarized light \citep{OMCDiS98}.
Additional indicators of the presense of an AGN include the existence of
a non-stellar source at 2.2$\,\mu$m whose diameter is less than 3\,pc
\citep[][though these authors see no compelling evidence for a nuclear
black hole]{MKTG98}, H$_2$O
maser activity \citep{GW82,Getal97}, Fe-K fluorescent line emission detected
in its X-ray spectrum \citep{MFPPFGMOS96}, and
observed high excitation lines of Ne, S, Mg, O, and Si in the 2.5--45$\,\mu$m
wavelength range \citep[][but see also \citealt{BWRS-B97} and \citealt{OMM99}]
{Moorwoodetal96}.
Recent star
formation in the nuclear region has been detected through the near-IR
observations of \citet{MKTG98}, who find that the luminosity due to starformation
within the central few hundred parsecs is comparable to the AGN luminosity,
though the central 14pc (converted to our adopted distance) is dominated by the
AGN luminosity, with starformation contributing only 2\% of the light.  Additional
evidence for a nuclear starburst is also provided by the morphology of extended
H$\alpha$ emission, produced as a result of a nuclear ``superwind'' \citep[][see also
\citealt{EHJSE98}]{LH96}.

\subsection{NGC\,5728}

This galaxy is a classic example of a Seyfert~2 with biconical emission line
cones \citep{schommer88, WBHKM93, CAMSB96a}, separated by a dark band
\citep{WBHKM93}.  These ionization cones are
essentially colinear, with the position angle of the NW cone axis being
304$^\circ$ and that of the brighter SE cone being 118$^\circ$.  The radio
continuum maps at 6 and 20\,cm of the nuclear region of NGC\,5728 show
extensions of similar length and at essentially the same position angles,
though the NW extension is the brighter of the two \citep{WBHKM93}.
HST imaging polarimetry of the central regions of NGC\,5728 has been used
to determine the position of the nucleus which lies behind the dark band
\citep{CAMSB96a}.  These authors place the nucleus at a point defined by the
apexes of the two ionization cones, in agreement with \citet{WBHKM93}.
Apart from the cones, the nuclear region of NGC\,5728 shows other interesting
features.  In the red and green optical continuum images presented by
\citet{WBHKM93}, there exist two miniature bars delineated by four main knots
of emission (A--D) and aligned EW (that is, colinear with neither the cones
or the larger scale stellar bar).  We note in this context that our
K-band image is consistent with this innermost bar feature, seen at poorer
spatial resolution.

\subsection{NGC\,7469}

NGC\,7469 is often cited as the prototypical Seyfert~1 galaxy.
A variety of observations at high spatial resolution indicate that the active
nucleus is surrounded by a more or less complete ring of starburst activity
\citep{mauder94,genzel95,MHH94,whhw91}. \citet{genzel95} have modelled their
high resolution NIR imaging and imaging spectroscopic data\footnote{The NIR
imaging spectroscopic observations of NGC\,7469 reported on by \citet{genzel95}
were made with MPE's NIR imaging Fabry-Perot FAST as well as 3D\@.  The 3D
observations were made during the 3D commissioning run and did not benefit
from tip-tilt guiding.  3D observations reported on in this work were made
on a subsequent run with tip-tilt correction.} and find that two thirds of the
bolometric luminosity of the entire galaxy originates in this starburst ring.
\citet{PC96} cannot spatially resolve the starburst ring in their {\em ROSAT\/}
HRI data but derive its X-ray luminosity by using the results of the
radio observations of \citet{whhw91} together with the assumption that its
$L_\mathrm{X}/L_{\mathrm{5\,GHz}}$ ratio is typical of starbursts ($\sim$400).
They then arrive at the conclusion that only 4\% of the total X-ray luminosity
is produced in the ring.
The active nucleus itself is seen to vary in the X-rays over timescales from
hours \citep{Barr86} to days \citep{LKAT96}.  \citet{DLR89} report on nuclear
optical variations with timescales ranging from hours to years.  In the
infrared, \citet{glass98} found small ($<$0.3\,mag) variations during the
course of 13~years of observations, except for a substantial dimming in 1989
during which the nuclear flux dropped to one fifth its normal value.  Related
decreases in the U and B brightnesses, along with the disappearance of the
broadline component of H$\beta$ are reported by
\citet[][and references therein]{glass98}.

\subsection{Mrk\,315 (=II\,Zw\,187)}

Mrk\,315, which has been classified as a Seyfert~1.5 \citep{koski78}, has
a ``jet-like'' structure which extends straight at position angle 324$^\circ$
for 45~$h^{-1}$~kpc before curving back towards the southeast
\citep{MacKenty86,msguw94}.  The origin for this feature, which is seen only
in [O{\small III}] and H$\alpha$ \citep[but see also][]{NHFFD98}, is likely
due to a tidal interaction.  A knot is seen at optical, near-infrared, and
radio wavelengths which lies 2$^{\prime\prime}$ to the east of the nucleus
\citep[][see also our K-band image]{MacKenty86,msguw94,NHFFD98}.  The nature
of this knot is not certain, with \citet{msguw94} describing it as a recently
captured galaxy remnant and \citet{NHFFD98} attributing it to an intense
starburst region.  An extended region of radio continuum emission asymmetrically
envelopes the nucleus and this secondary knot \citep{UWS81, UW84, NHFFD98}
and is consistent with a starburst origin \citep{Wilson88}.  Within this radio
continuum region lies two complex arms or arcs of emission.  The brighter of
the two appears to connect the nucleus and the secondary peak, and then arc
northward of the nucleus
\citep[especially well shown in Figure~4 of][]{NHFFD98}.  The second arc lies
to the west of the nucleus and is blue in color, appearing in J- and H-band
images but not at K-band \cite{NHFFD98}.  These arcs are likely due to the
recent tidal interaction used to explain the secondary peak \citep{msguw94}.

\subsection{NGC\,7582}

NGC\,7582 presents a fascinating puzzle.  It is a classical Seyfert~2
galaxy, with a well-defined [O{\small III}] cone \citep{Morrisetal85,
S-BB91} and normally narrow emission lines \citep[e.g.][]{CFS-BS98}.
Its lack of broad lines in spectropolarimetric observations coupled with
a high 60$\,\mu$m/25$\,\mu$m flux ratio led
\citet{HLB97} to infer that the nucleus itself is totally obscured, even
to scattered light, by an edge-on thick torus surrounding it.
This is also consistent with the results of {\em Ginga\/} observations
of NGC\,7582 \citep{Warwicketal93}.
That stars contribute to the nuclear luminosity of NGC\,7582 has been
demonstrated by the work of \citet{OOKM95}.  They find, through observations
of strong CO bandhead absorption lines and an inferred large near-IR
light-to-mass ratio that young supergiants are present and dominate
the H- and K-band nuclear luminosity \citep[but see also][who find, based on
spectral synthesis fitting to optical spectra, that only 6\% of the optical
light is due to young stars]{SS-BCF99}.  There is also a kiloparsec-scale
disk of H{\small II} regions surrounding the nucleus \citep{Morrisetal85}.

Recently, \citet{AJKMT99} report on the detection of hitherto unseen broad
lines in the optical spectra, effectively for a few months turning this
classical Seyfert~2 galaxy into a Seyfert~1 galaxy.  These authors debate
the pros and cons of three models to explain the sudden appearance of the
broad lines: the capture of a star by a nuclear black hole, a reddening change
in the surrounding obscuring torus, and the radiative onset of a Type~IIn
supernova.  The first two of these models deal directly with the active
nucleus and its obscuring surroundings, while the last model invokes pure
stellar processes.  \citet{AJKMT99} raise serious concerns with the first two
models, and favor the SN theory.  Evidence in favor of a nuclear-based
explanation comes from the {\em BeppoSAX\/} observations of \citet{TPFMGPB00}.
They observed NGC\,7582 in 1998~November and found a
previously unseen hard X-ray component.  Correlated variability across the
X-ray spectrum led \citet{TPFMGPB00} to the conclusion that a single
component dominates the 2--100\,keV band.  Further, these variations do not
seem to be at all correlated with those in the optical reported on by
\citet{AJKMT99}.

NGC\,7582 was observed twice in the K-band, once before and once just after the
optical anomaly.  The second set of observations were done after the anomaly
was brought to our attention (R.~Terlevich, private communication). A
more detailed analysis of our observations will be presented in a subsequent
paper, but here we note simply that our second observations show a
broadening of the Br$\gamma$ line, along with a reddening of the continuum
in the seeing
weighted aperture.  Our observed Br$\gamma$ equivalent width for the 1994 July
ESO 2.2m observations agrees well with that obtained by \citet{OOKM95} who
also observed NGC\,7582 during its ``quiescence.''

\acknowledgments
We thank the staffs at Siding Spring, Calar Alto, La~Palma, and ESO La~Silla
for their expert support during these observations.  Many thanks also go
to all of the (past and present) members of the MPE IR/Submillimeter
group who helped in these observations.  In particular, we would like to
single out Stephan~Anders, Alfred~Krabbe, Roberto~Maiolino,
Eva~Schinnerer, and Niranjan~Thatte for making reduced data sets of some
of the objects discussed in this paper available to us.  Thanks are also
extended to the referee, Hajime Sugai, for helpful suggestions.  This research
has made use of the NASA/IPAC Extragalactic Database (NED) which is
operated by the Jet Propulsion Laboratory, California Institute of
Technology, under contract with the National Aeronautics and Space
Administration.

\clearpage

% FIGURE 1

\begin{figure}[h]
%\epsscale{0.60}
%\plotone{fig1.eps}
\caption{Most relevant diagnostic diagram used for spectral classification.
Dots show the location of the sources in this paper. The \citet{BPT81} model
for H{\small II} regions is over-plotted, as well as the limits
[O{\small III}]$\lambda$5007/H$\beta = 3$ and
[N{\small II}]$\lambda$6583/H$\alpha = 0.5$. Three different regions are
delineated, the Seyfert region on the upper
right, the LINERs on the lower right and the H{\small II} around the plotted
model (about the dashed region).}
\end{figure}

% FIGURE 2

\begin{figure}[ht]
%\epsscale{0.87}
%\plotone{fig2a.eps}
\caption{H- and K-band spectra of the galaxies in the sample. 
Wavelengths are in the observed frame. The top
spectrum in each panel is the seeing-weighted (nuclear) spectrum (see text for
details).  The middle spectrum in each panel shows a spectrum extracted over a
3\arcsec\ or 4\arcsec\ (as indicated in the upper right-hand
corner of each plot) uniformally weighted aperture, while the bottom spectrum
of each panel (labeled as 'sub') shows the difference between
the middle and top spectra.  In the
bottom box for each panel,
prominent stellar absorption lines (even if not detected) and all of the
significantly detected emission lines are marked. The $\earth$ symbol indicates telluric features. }
\end{figure}

%\clearpage
%\begin{figure}[ht]
%\epsscale{0.87}
%\plotone{fig2b.eps}
%\figurenum{2}
%\caption{continued}
%\end{figure}

%\clearpage
%\begin{figure}[ht]
%\epsscale{0.87}
%\plotone{fig2c.eps}
%\figurenum{2}
%\caption{continued}
%\end{figure}

%%\clearpage
%\begin{figure}[ht]
%\epsscale{0.87}
%\plotone{fig2d.eps}
%\figurenum{2}
%\caption{continued}
%\end{figure}

%\clearpage
%\begin{figure}[ht]
%\epsscale{0.87}
%\plotone{fig2e.eps}
%\figurenum{2}
%\caption{continued}
%\end{figure}

%\clearpage
%\begin{figure}[ht]
%\epsscale{0.87}
%\plotone{fig2f.eps}
%\figurenum{2}
%\caption{continued}
%\end{figure}

%\clearpage
%\begin{figure}[ht]
%\epsscale{0.87}
%\plotone{fig2g.eps}
%\figurenum{2}
%\caption{continued}
%\end{figure}

%\clearpage
%\begin{figure}[ht]
%\epsscale{0.87}
%\plotone{fig2h.eps}
%\figurenum{2}
%\caption{continued}
%\end{figure}

%\clearpage
%\begin{figure}[ht]
%\epsscale{0.87}
%\plotone{fig2i.eps}
%\figurenum{2}
%\caption{continued}
%\end{figure}

%\clearpage
%\begin{figure}[ht]
%\epsscale{0.87}
%\plotone{fig2j.eps}
%\figurenum{2}
%\caption{continued}
%\end{figure}

%\clearpage
%\begin{figure}[ht]
%\epsscale{0.87}
%\plotone{fig2k.eps}
%\figurenum{2}
%\caption{continued}
%\end{figure}

%\clearpage
%\begin{figure}[ht]
%\epsscale{0.87}
%\plotone{fig2l.eps}
%\figurenum{2}
%\caption{continued}
%\end{figure}

%\clearpage
%\begin{figure}[ht]
%\epsscale{0.87}
%\plotone{fig2m.eps}
%\figurenum{2}
%\caption{continued}
%\end{figure}

% FIGURE 3

\begin{figure}[ht]
%\epsscale{1.000}
%\plotone{fig3a.eps}
\caption{
Images from the 3-dimensional data cube of all the galaxies in the sample.
The object names and
bands over which the images were generated are indicated in the label above each
plot and the axes show the offset relative to the broad band peak for each
galaxy in each image.  In each map the contours start at 90\% of the peak
surface brightness listed in Tables~6--8, and continue as 80\%, 70\%, ...,
10\%, 5\%\@.
In all images, north is at the top, east is to the left.}
\end{figure}

\begin{deluxetable}{lccrcrcc}
\tablecolumns{8}
\tablewidth{0pt}
\tabletypesize{\footnotesize}
\tablenum{1}
\tablecaption{Observed Galaxy Sample and Adopted Distances}
\tablehead{
\colhead{Galaxy}& \colhead{RA}&
\colhead{Dec}&\colhead{Ref.}&\colhead{Galaxy Type}&
\colhead{$V_{3K}$}&\colhead{Distance (Mpc)}&\colhead{Note} \\
\colhead{(1)}&\colhead{(2)}&
\colhead{(3)}&\colhead{(4)}&
\colhead{(5)}&\colhead{(6)}&
\colhead{(7)}&\colhead{(8)}}
\startdata
Mrk\,348   & 00 48 47.1 & $+$31 57 25 &  1 & SA(s)0/a:        &  4411 &  59   & a \\
I\,Zw\,1   & 00 53 34.9 & $+$12 41 36 &  2 & S?               & 17870 & 238   & a \\
Mrk\,573   & 01 43 57.8 & $+$02 21 00 &  2 & (R)SAB(rs)0$^+$: &  4871 &  65   & a \\
Mrk\,1044  & 02 30 05.4 & $-$08 59 53 &  3 & S?               &  4660 &  62   & a \\
NGC\,1068  & 02 42 40.7 & $-$00 00 48 &  4 & (R)SA(rs)b       &   871 & 14.4  & b \\
NGC\,1097  & 02 46 19.1 & $-$30 16 28 &  5 & SB(s)b           &  1110 & 14.5  & b \\
NGC\,1194  & 03 03 49.2 & $-$01 06 13 &  6 & SA0$^+$:         &  3764 &  50   & a \\
NGC\,1275  & 03 19 48.1 & $+$41 30 42 &  7 & Pec              &  5100 &  68   & a \\
NGC\,1365  & 03 33 36.4 & $-$36 08 25 &  8 & SB(s)b           &  1579 &  18.6 & c \\
NGC\,1386  & 03 36 45.4 & $-$35 59 57 &  9 & SB(s)0$^+$       &   772 &  18.6 & d \\
NGC\,1433  & 03 42 01.4 & $-$47 13 20 & 10 & (R')SB(r)ab      &   911 &  11.6 & b \\
NGC\,1566  & 04 20 00.6 & $-$54 56 17 &  9 & SAB(s)bc         &  1440 &  13.4 & b \\
NGC\,1672  & 04 45 42.1 & $-$59 14 57 & 11 & SB(s)b           &  1300 &  14.5 & b \\
Mrk\,1095  & 05 16 11.4 & $-$00 08 59 &  2 & S?               &  9936 & 132   & a \\
NGC\,2110  & 05 52 11.4 & $-$07 27 22 & 12 & SAB0$^-$         &  2353 &  31   & a \\
NGC\,3079  & 10 01 57.8 & $+$55 40 47 & 13 & SB(s)c sp        &  1260 &  20.4 & b \\
NGC\,3227  & 10 23 30.6 & $+$19 51 54 &  2 & SAB(s)a pec      &  1466 &  20.6 & b \\
NGC\,3393  & 10 48 23.4 & $-$25 09 43 & 14 & (R')SB(rs)a:     &  4079 &  54   & a \\
NGC\,3783  & 11 39 01.8 & $-$37 44 19 &  9 & (R')SB(r)ab      &  3242 &  43   & a \\
NGC\,4051  & 12 03 09.6 & $+$44 31 53 &  2 & SAB(rs)bc        &   910 &  17.0 & b \\
NGC\,4258  & 12 18 57.5 & $+$47 18 14 & 15 & SAB(s)bc         &   684 &   7.3 & e \\
NGC\,4639  & 12 42 52.4 & $+$13 15 27 & 16 & SAB(rs)bc        &  1216 &  25.5 & f \\
Mrk\,231   & 12 56 14.2 & $+$56 52 25 & 17 & SA(rs)c pec      & 12440 & 166   & a \\
NGC\,4945  & 13 05 27.5 & $-$49 28 06 & 18 & SB(s)cd sp       &   826 &   3.7 & g \\
Mrk\,273   & 13 44 42.1 & $+$55 53 13 & 12 & Pec              & 11400 & 152   & a \\
Mrk\,463W  & 13 56 02.6 & $+$18 22 18 &  9 & S?               & 15537 & 205   & a \\
Mrk\,463E  & 13 56 02.9 & $+$18 22 19 &  2 & S?               & 15388 & 205   & a \\
Circinus   & 14 13 09.3 & $-$65 20 21 & 19 & SA(s)b:          &   530 &  4.2  & h \\
NGC\,5728  & 14 42 23.9 & $-$17 15 11 &  9 & SAB(r)a:         &  3111 &  41   & a \\
NGC\,7469  & 23 03 15.6 & $+$08 52 26 &  2 & (R')SAB(rs)a     &  4477 &  60   & a \\
Mrk\,315   & 23 04 02.6 & $+$22 37 28 &  2 & E1 pec           & 11280 & 150   & a \\
NGC\,7582  & 23 18 23.5 & $-$42 22 14 & 20 & (R')SB(s)ab      & 1325  & 17.6  & b
\enddata
\end{deluxetable}

\clearpage

Note. ---
Col.~(1)         --- Source designation.
Col.~(2) and (3) --- Right Ascension and declination (J2000).
Col.~(4)         --- Reference for the coordinates listed in Col.~(2) and (3).
                      1) \citet{Wilkinsonetal98},  2) \citet{Clements81},
                      3) \citet{ArgEld90},         4) \citet{Capettietal97},
                      5) \citet{Cozioletal94},     6) \citet{Falcoetal99},
                      7) \citet{Johnstonetal95},   8) \citet{Lindbladetal96},
                      9) \citet{activecat},       10) \citet{Maozetal96},
                     11) \citet{Lauberts82},      12) \citet{Clements83},
                     13) \citet{BaaIrw95},        14) \citet{Faundez98},
                     15) \citet{TurnerHo94},      16) \citet{2Mass},
                     17) \citet{Maetal98},        18) \citet{GMH97},
                     19) \citet{Fairalletal98},   20) \citet{Joguetetal98}
Col.~(5)         --- Galaxy classification of the host as given by \citet{RC3}.
                     Note that \citet{RC3} make no distinction between
                     Mkn\,463W and Mkn\,463E.
Col.~(6)         --- Optical systemic velocity, $V_{obs}$, (in km~s$^{-1}$) from \citet{RC3}
                     (except for NGC\,1194 [taken from \citet{gkmgl92}],
                     Mkn\,463W [taken from \citet{activecat}], and Mkn\,463E
                     [taken from \citet{HewBur91}), corrected to the microwave
                     background frame, using Eq.~(1).
Col.~(7)         --- Distance to the source, in Mpc.
Col.~(8)         --- The source for the adopted distance:
             a) $\mathrm{D}=V_{3K}/\mathrm{H}_0$, where H$_0$ is taken to be
                75\,km~sec$^{-1}$~Mpc$^{-1}$,
             b) taken from \citet{T88} which corrects for Virgocentric infall
                of 300\,km~sec$^{-1}$,
             c) HST observations of Cepheid variables \citep{metal99},
             d) same value as NGC\,1365 (see Note c) as they are in the same group,
             e) water maser measurements \citep{Hetal97},
             f) HST observations of Cepheid variables \citep{SSLTMP97},
             g) value based on the mean distance modulus to
                NGC\,5128 \citep{ferrarese00}, since NGC\,4945 is in the same group,
             h) average value from \citet{Fetal77}, corrected to our adopted
                H$_0$ value.

\clearpage

\begin{deluxetable}{ccccccccc}
\tablecolumns{9}
\tablewidth{0pt}
\tabletypesize{\footnotesize}
\tablenum{2}
\tablecaption{Summary of Optical Emission Ratios and Seyfert Class}
\tablehead{
\colhead{Galaxy}&\colhead{[O{\scriptsize III}]/H$\beta$}&
\colhead{[N{\scriptsize II}]/H$\alpha$}&\colhead{[S{\scriptsize II}]/H$\alpha$}&
\colhead{[O{\scriptsize II}]/[O{\scriptsize III}]}&\colhead{[O{\scriptsize I}]/H$\alpha$}&
\colhead{[O{\scriptsize I}]/[O{\scriptsize III}]}&\colhead{Class}&
\colhead{Refs}\\
\colhead{(1)}&\colhead{(2)}&
\colhead{(3)}&\colhead{(4)}&
\colhead{(5)}&\colhead{(6)}&
\colhead{(7)}&\colhead{(8)}&
\colhead{(9)}}
\startdata
Mrk\,348  &   11.74 &    0.83  &    0.85  &    0.38  &    0.39  &    0.091 & S2                     & 1    \\
I\,Zw\,1  &    0.82 & \nodata  & \nodata  &    0.48  & \nodata  & \nodata  & NLS1/H{\scriptsize II} & 2    \\
Mrk\,573  &   12.3  &    0.84  &    0.53  &    0.19  &    0.10  &    0.026 & S2                     & 3    \\
Mrk\,1044 &    1.67 & \nodata  &    0.06  & \nodata  &    0.008 &    0.015 & NLS1                   & 4    \\
NGC\,1068 &   13.14 &    0.82  &    0.09  &    0.035 &    0.10  &    0.031 & S2                     & 5    \\
NGC\,1097 &    5.1  &    3     &    0.7   &    0.78  & \nodata  & \nodata  & S1.5                   & 6    \\
NGC\,1194 & $>$8.42 &    0.29  & \nodata  & \nodata  & \nodata  & \nodata  & S1.5                   & 7    \\
NGC\,1275 &    3.63 &    0.42  &    0.3   &    0.99  &    0.71  &    0.42  & S1.5/L                 & 8    \\
NGC\,1365 &    3    &    0.55  &    0.34  &    0.92  & \nodata  & \nodata  & S1.5/H{\scriptsize II} & 9,10 \\
NGC\,1386 &   12.2  &    1.276 &    0.56  &    0.178 &    0.13  &    0.032 & S2                     & 11   \\
NGC\,1433 &    2    &    1.54  &    0.98  &    1.15  & \nodata  & \nodata  & L                      & 9,12 \\
NGC\,1566 &    6.5  &    0.35  &    0.18  &    1.04  & \nodata  & \nodata  & S1                     & 13   \\
NGC\,1672 &    2.1  &    1.41  &    0.4   &    0.62  & \nodata  & \nodata  & L                   & 6    \\
Mrk\,1095 &    5.6  &    0.8   &    0.48  & \nodata  & $<$0.28  & $<$0.12  & S1                     & 4    \\
NGC\,2110 &    4.76 &    1.41  &    1.11  &    0.92  &    0.45  &    0.25  & S2                     & 2,14 \\
NGC\,3079 &    4.15 &    1.59  &    0.86  & \nodata  &    0.18  &    0.13  & S2                     & 15   \\
NGC\,3227 &    5.91 &    1.33  &    0.68  &    0.177 &    0.23  &    0.12  & S1.5                   & 15,16\\
NGC\,3393 &   10.2  &    1.2   & \nodata  &    0.22  &    0.12  &    0.034 & S2                     & 17,18   \\
NGC\,3783 &    5.5  &    0.18  &    0.14  &    0.076 &    0.081 &    0.043 & S1                     & 19   \\
NGC\,4051 &    4.5  &    0.64  &    0.36  &    0.292 &    0.14  &    0.093 & S1                     & 15,16\\
NGC\,4258 &   10.32 &    0.80  &    0.94  & \nodata  &    0.38  &    0.11  & S1.5                   & 15   \\
NGC\,4639 &    3.77 &    1.12  &    1.09  & \nodata  &    0.31  &    0.25  & S1                     & 15   \\
Mrk\,231  & \nodata & \nodata  & \nodata  & \nodata  & \nodata  & \nodata  & S1                     & 20   \\
NGC\,4945 & \nodata &    1.3   &    0.9   & \nodata  & \nodata  & \nodata  & H{\scriptsize II}/S2/L & 21   \\
Mrk\,273  &    3.63 &    0.87  &    0.50  &    2.21  &    0.081 &    0.063 & S2/L                   & 24   \\
Mrk\,463W &    3.0  &    0.57  &    0.62  & \nodata  & \nodata  & \nodata  & S2                     & 23   \\
Mrk\,463E &    8.5  &    0.45  &    0.39  & \nodata  &    0.14  &    0.040 & S2                     & 23   \\
Circinus  &   10.25 &    1.15  &    0.64  &    0.26  &    0.14  &    0.41  & S2                     & 25   \\
NGC\,5728 &   11.8  &    1.40  &    0.33  &    0.32  &    0.17  &    0.085 & S2                     & 2,22 \\
NGC\,7469 &    6.07 &    0.59  &    0.28  &    0.40  &    0.077 &    0.079 & S1                     & 19   \\
Mrk\,315  &    2.56 &    0.51  &    0.22  &    0.86  &    0.049 &    0.062 & S1.5/H{\scriptsize II} & 1    \\
NGC\,7582 &    2.33 &    0.69  &    0.274 &    0.482 &    0.027 &    0.042 & S2/H{\scriptsize II}   & 19
\tablecomments{Col.~(1) --- Source designation.  Col.~(2) through (7)
--- Nuclear line ratios of our Seyfert sample.  All but I\,Zw\,1, Mrk\,1044,
NGC\,1194, NGC\,4945, and NGC\,7469 were corrected for extinction.
An underlying absorption-correction was made for all sources except I\,Zw\,1,
Mrk\,1044, NGC\,1068, and Mrk\,1095.
For NGC\,1365 the listed
[N{\scriptsize II}]/H$\alpha$ and [S{\scriptsize II}]/H$\alpha$ ratios
correspond to a position
2\arcsec\ to the south of the nucleus.
Col.~(8) --- AGN classification (see text for details).
Col.~(9) --- References for emission line properties:
1)~\citet{koski78}, 2)~\citet{ccsgdb94}, 3)~\citet{swmb96},
4)~\citet{RS91}, 5)~\citet{KRC98}, 6)~\citet{SWB96}, 7)~\citet{gkmgl92},
8)~\citet{HFS93}, 9)~\citet{AK79}, 10)~\citet{SKSM99},
11)~\citet{srswb96}, 12)~\citet{SKC95}, 13)~\citet{FK89},
14)~\citet{S80}, 15)~\citet{HFS97}, 16)~\citet{Schmitt98},
17)~\citet{ferguson97}, 18)~\citet{CBFNW2000}, 19)~\citet{BBA89}, 20)~\citet{sanders88},
21)~\citet{LTM97}, 22)~\citet{PCB83}, 23)~\citet{CV95},
24)~\citet{LK95}, 25)~\citet{OMM99}.  See text for details on Mrk\,231 and NGC\,4945.}

\enddata
\end{deluxetable}

\clearpage

\pagebreak

\begin{deluxetable}{ccc}
\tablecolumns{5}
\tablewidth{0pt}
\tablenum{3}
\tablecaption{Definition of Observed Bands}
\tablehead{
\colhead{Band}&\colhead{Bandwidth}&
\colhead{$R$} \\
\colhead{(1)}&\colhead{(2)}&
\colhead{(3)}}
\startdata
H  & 1.48--1.78 & 1250 \\
HH & 1.55--1.75 & 2100 \\
K  & 1.94--2.41 & 1100 \\
KL & 2.17--2.43 & 2100
\enddata
\tablecomments{Col.~(1) --- Band designation.  Col.~(2) --- Wavelength coverage
in microns.  Col.~(3) --- Resolution, $R=\Delta\lambda/\lambda$}
\end{deluxetable}

\clearpage

\begin{deluxetable}{llccl}
\tablecolumns{5}
\tablewidth{0pt}
\tablenum{4}
\tablecaption{Observing Runs}
\tablehead{
\colhead{Observatory}&\colhead{Telescope}&
\colhead{pixel size}&\colhead{FOV (\arcsec)}&
\colhead{Dates of Runs} \\
\colhead{(1)}&\colhead{(2)}&
\colhead{(3)}&\colhead{(4)}&
\colhead{(5)}}
\startdata
Calar Alto     & 3.5m     &  0.3, 0.5 & 4.8, 8.0 & Nov/Dec 1993, Jan 1995, Jul/Aug 1997, Jul 1998 \\
La Palma       & WHT/4.2m &  0.3, 0.5 & 4.8, 8.0 & Apr 1994, Dec 1995/Jan 1996\\
ESO            & 2.2m     &  0.3, 0.5 & 4.8, 8.0 & Jul/Aug 1994, Mar 1996, Apr 1996 \\
Siding Springs & AAT/3.9m & 0.25, 0.4 & 4.0, 6.4 & Nov 1997, Dec 1997, Feb 1998, Mar 1998 \\
               &          &           &          & Oct 1998, Nov/Dec 1998, Feb 1999, Apr 1999
\enddata
\end{deluxetable}

\clearpage

\begin{deluxetable}{lrcccccccc}
\tablecolumns{10}
\tablewidth{0pt}
\tabletypesize{\scriptsize}
\tablenum{5}
\tablecaption{Summary of Data Obtained and Literature Photometry}
\tablehead{
\colhead{Galaxy}&\colhead{Int}&
\colhead{Obs}&\colhead{Date}&\colhead{Band}&
\colhead{seeing}&
\colhead{Pixel Scale}&
\colhead{K}&
\colhead{H}&\colhead{Aperture} \\
\colhead{(1)}&\colhead{(2)}&
\colhead{(3)}&\colhead{(4)}&
\colhead{(5)}&\colhead{(6)}&
\colhead{(7)}&\colhead{(8)}&
\colhead{(9)}&\colhead{(10)}}
\startdata
Mrk\,348  &    1200 &    3.5m &    8/97 &      K  & 1.0      & 0.5     & 12.17  &  12.72  &     3   \\
I\,Zw\,1  &    4200 &    3.5m &    1/95 &      K  & 1.0      & 0.5     &  9.90  &  11.00  &     8.5 \\
          &    2370 &    WHT  &   12/95 &      H  & 1.0--1.5 & 0.5     &\nodata & \nodata & \nodata \\
Mrk\,573  &    4480 &    2.2m &    8/94 &      K  & 1.5      & 0.5     & 11.78  &  12.27  &     3   \\
Mrk\,1044 &    8760 &    2.2m &    8/94 &      K  & 1.5      & 0.5     &\nodata & \nodata & \nodata \\
          &    3300 &    AAT  &   10/98 &      HH & 1.5      & 0.4     &\nodata & \nodata & \nodata \\
NGC\,1068 &    5560 &    WHT  &    1/96 &      K  & 0.7      & 0.3     &  7.65  &   8.53  &     6   \\
          &    1800 &    AAT  &   11/97 &      H  & 1.0      & 0.4     &\nodata & \nodata & \nodata \\
NGC\,1097 &    1200 &    AAT  &   11/97 &      K  & 0.9      & 0.4     &  9.90  &  10.22  &     6   \\
          &    1800 &    AAT  &   11/97 &      H  & 0.9      & 0.4     &\nodata & \nodata & \nodata \\
NGC\,1194 &    2000 &    AAT  &   12/98 &      K  & 2.0      & 0.4     &\nodata & \nodata & \nodata \\
NGC\,1275 &    4680 &    3.5m &    1/95 &      K  & 1.0--1.5 & 0.5     & 11.86  &  12.82  &     3   \\
          &    1920 &    3.5m &    1/95 &      H  & 1.3      & 0.5     &\nodata & \nodata & \nodata \\
NGC\,1365 &    3000 &    AAT  &   12/97 &      K  & 0.7      & 0.25    &  9.00  &   9.73  &     9.15\\
          &    2520 &    AAT  &   12/97 &      H  & 0.5--1.0 & 0.4     &\nodata & \nodata & \nodata \\
NGC\,1386 &    2800 &    AAT  &   12/97 &      K  & 1.0      & 0.4     & 10.12  &  10.45  &     6   \\
          &    2400 &    AAT  &   12/97 &      H  & 0.8      & 0.4     &\nodata & \nodata & \nodata \\
NGC\,1433 &    2600 &    AAT  &   12/97 &      K  & 0.9      & 0.4     &  9.09  &   9.35  &    18   \\
          &    2760 &    AAT  &   12/97 &      H  & 0.8--1.6 & 0.4     &\nodata & \nodata & \nodata \\
NGC\,1566 &    2600 &    AAT  &   11/97 &      K  & 0.5      & 0.4     &  9.85  &  10.36  &     6   \\
          &    2880 &    AAT  &   11/97 &      H  & 0.8      & 0.4     &\nodata & \nodata & \nodata \\
NGC\,1672 &    3200 &    AAT  &   12/97 &      K  & 0.9      & 0.4     & 10.73  &  11.31  &     3   \\
          &    1680 &    AAT  &   12/97 &      H  & 0.5      & 0.25    &\nodata & \nodata & \nodata\\
Mrk\,1095 &    3360 &    WHT  &   12/95 &      K  & 1.2      & 0.5     & 10.39  &  11.26  &     4.6 \\
          &    2700 &    WHT  &    1/96 &      H  & 1.2      & 0.3     &\nodata & \nodata & \nodata \\
NGC\,2110 &     340 &    WHT  &    1/96 &      K  & 0.5      & 0.3     & 10.77  &  11.41  &     3   \\
          &    2655 &    WHT  &    1/96 &      H  & 1.7      & 0.5     &\nodata & \nodata & \nodata \\
NGC\,3079 &    1530 &    3.5m &   12/93 &      K  & 2.0      & 0.5     & 10.49  &  11.24  &     3   \\
NGC\,3227 &    4000 &    WHT  &    1/96 &      K  & 1.0      & 0.5     &  9.85  &  10.35  &     8.5 \\
          &    1320 &    WHT  &    1/96 &      H  & 1.0      & 0.5     &\nodata & \nodata & \nodata \\
NGC\,3393 &    6400 &    AAT  &    3/98 &      KL & 0.8--1.4 & 0.4     & 11.51  &  11.8   &     3   \\
NGC\,3783 &    1800 &    AAT  &   12/97 &      K  & 1.5      & 0.4     &  9.88  &  10.82  &     6   \\
          &    1440 &    AAT  &   12/97 &      H  & 1.3      & 0.4     &\nodata & \nodata & \nodata \\
NGC\,4051 &    3120 &    WHT   &  12/95 &      K  & 0.9      & 0.3     & 10.05  &  10.86  &     8.5 \\
          &    2130 &    WHT  &   12/95 &      H  & 1.0      & 0.3     &\nodata & \nodata & \nodata \\
NGC\,4258 &    5200 &    3.5m &    1/95 &      K  & 1.0      & 0.5     &  9.74  &  10.03  &     7.2 \\
NGC\,4639 &    3600 &    AAT  &    3/98 &      KL & 1.4      & 0.4     &\nodata & \nodata & \nodata \\
Mrk\,231  &    3580 &    WHT  &    1/96 &      K  & 1.0      & 0.5     &  8.97  &  10.37  &     8.5 \\
NGC\,4945 &    1680 &    AAT  &    4/99 &      K  & 2.3      & 0.4     &  9.32  &  10.60  &     6   \\
          &    1320 &    AAT  &    4/99 &      HH & 1.6      & 0.4     &\nodata & \nodata & \nodata \\
Mrk\,273  &    5000 &    3.5m &    8/97 &      K  & 0.9      & 0.3     & 11.38  &  12.14  &     8.5 \\
Mrk\,463$^\ddag$  &    1420 &    WHT  &    4/94 &      K  & 1.5      & 0.5     & 10.86E &  12.54E &     2   \\
          & \nodata & \nodata & \nodata & \nodata &  \nodata & \nodata & 13.77W &  14.23W & \nodata \\
Circinus  &    8430 &    2.2m &    3/96 &      K  & 0.6      & 0.3     &  8.53  &   9.25  &     5   \\
          &   12200 &    2.2m &    4/96 &      KL & 0.8      & 0.3     &\nodata & \nodata & \nodata \\
NGC\,5728 &    8000 &    2.2m &    7/94 &      K  & 1.0--1.5 & 0.5     & 10.30  &  10.53  &     9.1 \\
NGC\,7469 &    6400 &    3.5m &    7/97 &      K  & 1.0      & 0.5     &  9.92  &  10.66  &     3.9 \\
          &    4040 &    3.5m &    7/97 &      HH & 1.0      & 0.3     &\nodata & \nodata & \nodata \\
Mrk\,315  &    8000 &    3.5m &    8/97 &      K  & 1.0      & 0.5     & 11.51  &  11.95  &     8.5 \\
NGC\,7582 &   11640 &    2.2m &    7/94 &      K  & 1.0--1.5 & 0.5     &  9.13  &   9.98  &     6   \\
          &    1600 &    AAT  &   10/98 &      K  & 1.0      & 0.4     &\nodata & \nodata & \nodata \\
          &    3400 &    AAT  &   10/98 &      HH & 1.1      & 0.4     &\nodata & \nodata & \nodata
\enddata
\end{deluxetable}

\clearpage

Note. ---
Col.~(1)         --- Source designation.
Col.~(2)         --- Total on source integration time (in seconds) over the bandpass
                     indicated in Col.~(5).
Col.~(3)         --- Telescope at which the observations were made.  See also
                     Table~4.
Col.~(4)         --- Date of the observations in the form month/year.
Col.~(5)         --- Band used.  See Table~3 for details.
Col.~(6)         --- Seeing during the observations in arc seconds. If the
                     overall variation in the seeing during the observations was
                     greater than 0.4\arcsec\ the range of seeing
                     during the total integration time is provided.
Col.~(7)        --- The projected pixel size in arcseconds for our
                     observations.
Col.~(8) and (9) --- K- and H-Band aperture magnitudes from \citet{deVL88} for
                     most of the galaxies, but also from \citet{a-hsww98} for
                     NGC\,1068, NGC\,3393, Mrk\,348, Mrk\,463 and Mrk\,573, and
                     from \citet{fwdbs92} for NGC\,1275, NGC\,1672, NGC\,2110
                     and NGC\,3079.
Col.~(10)         --- Aperture in arcseconds through which the magnitudes in
                     Cols.~(8) and (9) were measured.
                     Apertures chosen were those which most nearly matched our observed
                     field-of-view.  When this was not possible, the smallest
                     aperture available is quoted.
$^\ddag$Observations of Mrk\,463W and Mrk\,463E (observed simultaneously due to their
small separation) were compromised by a H-band leak in the K-band order selecting filter,
which was only discovered after the fact (see also \citealt{KCTK97}).

\clearpage

\begin{deluxetable}{cccccccccccc}
\tablecolumns{10}
\tablewidth{0pt}
\tabletypesize{\scriptsize}
\tablenum{6}
\tablecaption{K-band Flux and Properties of the Br$\gamma$ Emission Line}
\tablehead{
\colhead{Galaxy}& \colhead{K-band Flux}&
\colhead{K-band Peak}&\colhead{Br$\gamma$ Peak}&
\colhead{EW(Br$\gamma$)}& \colhead{$\sigma_{EW}$}&
\colhead{Flux}& \colhead{$\sigma_{Flux}$}&
\colhead{cz}& \colhead{$\sigma_{cz}$}&
\colhead{W}&\colhead{$\sigma_{W}$} \\
\colhead{(1)}&\colhead{(2)}&
\colhead{(3)}&\colhead{(4)}&
\colhead{(5)}&\colhead{(6)}&
\colhead{(7)}&\colhead{(8)}&
\colhead{(9)}&\colhead{(10)}&
\colhead{(11)}&\colhead{(12)}}
\startdata
Mrk\,348     &  0.17 &  0.24      & \nodata & 2.0     & 0.5     &  0.33    & 0.09    & 4370    & 70      & 530     & 110  \\
 2           &  0.37 &            &         & 2.3     & 0.7     &  0.83    & 0.27    & 4280    & 110     & 820     & 110  \\
 4           &  0.61 &            &         & $<$1.5  & \nodata &  $<$0.94 & \nodata & \nodata & \nodata & \nodata & \nodata\\
I\,Zw\,1     &  2.0  &  3.5       & 20      &    8.4  &    0.4  &    16    &  0.8    &   18050 &  70     &    2340 &  90  \\
 2           &  3.6  &            &         &    7.8  &    0.4  &    28    &   1.6   &   17980 &  80     &    2360 &   110  \\
 4           &  4.5  &            &         &    7.2  &    0.5  &    32    &  2.4    &   17970 &  110    &    2350 &   140 \\
Mrk\,573     &  1.1  &  0.78      & \nodata & 5.0     & 0.4     &  5.6     &  0.5    &  5300   &  25     &    540  &   50 \\
 2           &  1.5  &            &         & 4.9     &    0.6  &  7.4     &  0.9    &  5280   &  40     &    510  &  80 \\
 4           &  3.4  &            &         & 5.2     &    0.6  &  18      &  2.1    &  5320   &  30     &    540  &   60  \\
Mrk\,1044*   &  0.67 & \nodata    & \nodata &   15    &    0.9  &     9.8  &   0.6   &    4800 &  80     &    2350 &   140 \\
 2           &  0.92 &            &         &   14    &    1.2  &    13    &   1.1   &    4730 &  100    &    2250 &   180  \\
 4           &  1.5  &            &         &    19   &    1.2  &  28.7    &   1.8   &    5010 &  100    &    3180 &   130 \\
NGC\,1068$^\dagger$    &  8.5  & 24         & 23      &    2.3  &  0.1    &    19    &    0.9  & 990     &  20     &    806  &   40  \\
 2           & 23    &            &         &    2.5  &    0.1  &    56    &    3.0  &   972   &  26     &    775  &   40  \\
 4           & 30    &            &         &    2.6  &    0.2  &    77    &    4.5  &   980   &  30     &    785  &   50  \\
NGC\,1097    &  0.35 &  0.48      & \nodata & $<$0.81 & \nodata &  $<$0.29 & \nodata & \nodata & \nodata & \nodata & \nodata\\
 2           &  1.0  &            &         & $<$0.77 & \nodata &  $<$0.82 & \nodata & \nodata & \nodata & \nodata & \nodata\\
 4           &  2.8  &            &         & $<$0.72 & \nodata &  $<$2.1  & \nodata & \nodata & \nodata & \nodata & \nodata\\
NGC\,1194*   &  1.1  & \nodata    & \nodata & $<$0.71 & \nodata &  $<$0.77 & \nodata & \nodata & \nodata & \nodata & \nodata\\
 2           &  0.97 &            &         & $<$1.1  & \nodata &  $<$1.1  & \nodata & \nodata & \nodata & \nodata & \nodata\\
 4           &  2.4  &            &         & $<$0.85 & \nodata &  $<$2.0  & \nodata & \nodata & \nodata & \nodata & \nodata\\
NGC\,1275    &  0.30 &  0.40      &  4.4    &   9.8   &    0.9  &     3.0  &    0.3  &    5080 &    40   &     850 &   70   \\
 2           &  0.48 &            &         &    9.6  &    0.8  &     4.6  &    0.4  &    5070 &    30   &     880 &   60  \\
 4           &  0.88 &            &         &    8.5  &    0.7  &     7.5  &    0.6  &    5070 &    50   &    990  &   70  \\
NGC\,1365$^\dagger$    &  1.5  &  4.3       & 33      &    10.3 &    0.4  &    16    &   0.6   &     670 &    65   &    2690 &   100  \\
 2           &  4.8  &            &         &    9.0  &    0.4  &    43    &    2.0  &     815 &    50   &    2360 &   100  \\
 3           &  6.2  &            &         &    8.8  &    0.5  &    55    &    2.9  &     800 &    70   &    2360 &   160  \\
NGC\,1386    &  0.59 &  0.50      & \nodata & 1.0     & 0.2     &  0.6     & 0.1     & 850     &   70    & 260     & 90     \\
 2           &  0.95 &            &         & 1.1     &    0.3  &  1.0     & 0.3     & 850     &   60    & \nodata & \nodata  \\
 4           &  2.5  &            &         & $<$0.72 & \nodata &  $<$1.8  & \nodata & \nodata & \nodata & \nodata & \nodata\\
NGC\,1433$^\dagger$    &  0.15 &  0.15      & \nodata & $<$0.92 & \nodata &  $<$0.14 & \nodata & \nodata & \nodata & \nodata & \nodata\\
 2           &  0.28 &            &         & $<$0.92 & \nodata &  $<$0.26 & \nodata & \nodata & \nodata & \nodata & \nodata\\
 4           &  0.74 &            &         & $<$0.73 & \nodata &  $<$0.56 & \nodata & \nodata & \nodata & \nodata & \nodata\\
NGC\,1566    &  0.36 &  1.1       &  6.1    & 5.6     &  0.6    &  2.0     &  0.2    &  1720   &   70   &   1690  &   100  \\
 2           &  1.5  &            &         & 4.7     &  0.4    &  7.4     &  0.7    &  1630   &   50    &   1670  &   80  \\
 4           &  3.4  &            &         & 4.0     &  0.7    &  14      &  2.5    &  1710   &   110   &   1760  &  110  \\
NGC\,1672    &  0.39 &  0.54      & \nodata & $<$0.46 & \nodata &  $<$0.18 & \nodata & \nodata & \nodata & \nodata & \nodata\\
 2           &  1.1  &            &         & $<$0.37 & \nodata &  $<$0.42 & \nodata & \nodata & \nodata & \nodata & \nodata\\
 4           &  2.9  &            &         & $<$0.41 & \nodata &  $<$1.2  & \nodata & \nodata & \nodata & \nodata & \nodata\\
Mrk\,1095    &  1.1  &  1.3       & 17      &   14    &    0.7  &    16    &     0.8 &   9910  &    120  &    4300 &   170 \\
 2           &  1.9  &            &         &   14    &    0.7  &    27    &     1.4 &   9930  &     110 &    4260 &   160 \\
 4           &  2.8  &            &         &   12    &    0.7  &    34    &     1.9 &   9940  &     120 &    4100 &   160 \\
NGC\,2110    &  0.25 &  1.1       & \nodata &  1.2    & 0.3     &  0.31    & 0.07    & 2405    & 40      &   180   & 50    \\
 2           &  1.4  &            &         &  1.4    & 0.4     &    2.0   &    0.5  &  2330   &     50  & 310     & 80   \\
 4           &  2.4  &            &         &  $<$1.2 & \nodata &  $<$3.0  & \nodata & \nodata & \nodata & \nodata & \nodata\\
NGC\,3079    &  1.6  &  0.65      & \nodata & $<$1.8  & \nodata &  $<$2.9  & \nodata & \nodata & \nodata & \nodata & \nodata\\
 2           &  1.4  &            &         & $<$2.5  & \nodata &  $<$3.7  & \nodata & \nodata & \nodata & \nodata & \nodata\\
 4           &  3.3  &            &         & $<$2.0  & \nodata & $<$6.7   & \nodata & \nodata & \nodata & \nodata & \nodata\\
NGC\,3227    &  1.3  &  1.7       & 14      &    6.6  &    0.7  &     8.5  &   1.0   &    1140 &    60   &    1940 &   290  \\
 2           &  2.9  &            &         &    5.5  &    0.4  &    16    &   1.1   &    1180 &    60   &    1510 &   80  \\
 4           &  4.8  &            &         &    4.2  &    0.6  &    20    &   3.0   &    1190 &    90   &    1470 &   130  \\
NGC\,3393    &  0.30 &  0.26      & \nodata &    2.8  &    0.6  &    0.85  &   0.2   &    3610 &    60   &    650  &   130  \\
 2           &  0.55 &            &         &    2.9  &    0.6  &    1.6   &   0.3   &    3660 &    60   &    590  &   70  \\
 4           &  1.3  &            &         &    3.4  &    0.4  &     4.6  &   0.5   &    3640 &    40   &    380  &   50  \\
NGC\,3783    &  1.5  &  1.2       &  9.7    &   13    &    0.6  &    20    &    0.9  &    2720 &    80   &    2670 &   110  \\
 2           &  2.1  &            &         &   15    &    0.6  &    31    &    1.3  &    2750 &    80   &    2820 &   100  \\
 4           &  3.7  &            &         &    9.4  &    0.7  &    35    &    2.5  &    2370 &    170  &    2610 &   170 \\
NGC\,4051$^\dagger$    &  0.95 &  2.0       & 14      &    7.3  &    0.7  &     6.9  &    0.6  &     580 &    70   &    1870 & 120    \\
 2           &  2.1  &            &         &    6.2  &    0.9  &    13    &    1.9  &     630 &    90   &    1870 &   160 \\
 4           &  3.0  &            &         &    4.0  &    0.6  &    12    &    1.8  &     780 &    75   &    1610 &   80  \\
NGC\,4258    &  0.70 &  0.93      & \nodata & $<$0.58 & \nodata &  $<$0.42 & \nodata & \nodata & \nodata & \nodata & \nodata\\
 2           &  1.6  &            &         & $<$0.61 & \nodata &  $<$1.0  & \nodata & \nodata & \nodata & \nodata & \nodata\\
 4           &  3.7  &            &         & $<$0.57 & \nodata & $<$2.2   & \nodata & \nodata & \nodata & \nodata & \nodata\\
NGC\,4639*   &  1.0  & \nodata    & \nodata & \nodata & \nodata &  \nodata & \nodata & \nodata & \nodata & \nodata & \nodata\\
 2           &  0.96 &            &         & \nodata & \nodata &  \nodata & \nodata & \nodata & \nodata & \nodata & \nodata\\
 4           &  2.6  &            &         & \nodata & \nodata &  \nodata & \nodata & \nodata & \nodata & \nodata & \nodata\\
Mrk\,231     &  4.1  &  7.0       & \nodata & 2.3     & 0.3     &  9.5     &   1.0   &   12510 &  70     &   1270  &  100   \\
 2           &  7.6  &            &         & 2.1     & 0.3     &   16     &   2.2   &   12500 &  80    &   1240  &  150   \\
 4           &  9.0  &            &         & 2.5     & 0.3     &   23     &   2.7   &   12770 &  150    &   1640  &  160   \\
NGC\,4945    &  2.5  &  0.61      &  4.9    &    7.0  &    0.3  &    17    &    0.6  &     464 &   12    &   352   &  29   \\
 2           &  1.6  &            &         &    6.5  &    0.3  &    10.7  &    0.5  &     474 &   11    &   350   &  34   \\
 4           &  4.7  &            &         &    7.3  &    0.3  &    34    &    1.3  &     462 &   10    &   340   &  27   \\
Mrk\,273$^\dagger$     & 0.073 &  0.10      & \nodata &   11    &    1.7  &     0.8  &   0.1   &   11080 &   50    &     530 &  70   \\
 2           &  0.22 &            &         &    9.3  &    1.1  &     2.0  &   0.2   &   11130 &   50    &     490 &  50   \\
 4           &  0.57 &            &         &    5.9  &    0.9  &     3.3  &   0.5   &   11160 &   30    &     380 &  60   \\
Mrk\,463W    &  0.14 & \nodata    & \nodata & \nodata & \nodata & \nodata  & \nodata & \nodata & \nodata & \nodata & \nodata\\
 2           &  0.17 &            &         & \nodata & \nodata & \nodata  & \nodata & \nodata & \nodata & \nodata & \nodata\\
 3           &  0.28 &            &         & \nodata & \nodata & \nodata  & \nodata & \nodata & \nodata & \nodata & \nodata\\
Mrk\,463E    &  1.2  &  1.0       & \nodata & \nodata & \nodata & \nodata  & \nodata & \nodata & \nodata & \nodata & \nodata\\
 2           &  1.6  &            &         & \nodata & \nodata & \nodata  & \nodata & \nodata & \nodata & \nodata & \nodata\\
 3           &  2.3  &            &         & \nodata & \nodata & \nodata  & \nodata & \nodata & \nodata & \nodata & \nodata\\
Circinus     &  1.7  & 12/5.8     & 22      &     3.5 &    0.2  &     5.7  &     0.4 &     430 &    15   &     270 &   20  \\
 2           &  6.6  &            &         &     3.5 &    0.2  &    23    &    1.1  &     428 &    15   &     185 &   20  \\
 3           & 10    &            &         &     3.3 &    0.3  &    33    &     2.6 &     432 &    25   &     190 &   15  \\
NGC\,5728$^\dagger$    &  0.19 &  0.13      & \nodata &   5.6   &   1.2   &    1.1   &  0.2    &    2870 &   60    &   510   &   110 \\
 2           &  0.34 &            &         &   5.0   &   1.0   &    1.7   &  0.3    &    2870 &   70    &    530  &   100 \\
 4           &  1.1  &            &         &   3.0   &   0.8   &    3.3   &  0.9    &    2870 &   90   &   390   &     110 \\
NGC\,7469    &  1.2  &  1.7       & 15      &    10.0 &    0.5  &    12    &   0.6   &    4650 &    40   &    2020 &    140 \\
 2           &  2.6  &            &         &    9.7  &    0.5  &    25    &    1.4  &    4620 &    50   &    1950 &    190 \\
 4           &  4.5  &            &         &    8.9  &    0.4  &    40    &    1.9  &    4675 &    30   &    1150 &   60  \\
Mrk\,315     & 0.095 &  0.15      & \nodata &  7.2    &  2.0    &    0.67  &    0.2  &   11310 &    150  &    1230 &   150 \\
 2           &  0.21 &            &         &  6.6    &    1.9  &    1.4   &    0.4  &   11230 &    160  &    1160 &   140 \\
 4           &  0.46 &            &         &  3.9    &    0.9  &    1.7   &    0.4  &   11520 &    110  &    380  &   80 \\
NGC\,7582(E) &  1.8  &  1.5       &  9.1    &  5.1    &    0.4  &    9.2   &    0.6  &   1550  &   60    &    1300 &   100 \\
 2           &  3.1  &            &         &  5.4    &    0.5  &    17    &    1.4  &   1550  &   50    &    1360 &   110 \\
 4           &  6.9  &            &         &  5.9    &    0.5  &    41    &    3.1  &   1500  &   50    &    1150 &   140 \\
NGC\,7582(A) &  2.1  &  3.5       & 18      &    7.7  &    0.4  &    16    &    0.9  &    1570 &    50   &    1870 &   130  \\
 2           &  4.3  &            &         &    7.2  &    0.6  &    31    &    2.7  &    1495 &    70   &    1650 &   190  \\
 4           &  7.4  &            &         &    6.7  &    0.5  &    50    &    3.3  &    1528 &    30   &     770 &   50

%\tablecomments{}
\enddata
\end{deluxetable}

\clearpage

Note.~---
Col.~(1) --- Source designation.  Each galaxy has three associated rows with the
             first corresponding to the seeing-weighted aperture spectrum, the
             second being to a uniform aperture 2 arcseconds in diameter, and
             the third corresponding to either 3 or 4 arcseconds uniform
             aperture.  Two sets of data are listed for NGC\,7582.  The first
             are results obtained from the data taken at the ESO 2.2m telescope
             and the second are results of the data taken at the AAT\@.
Col.~(2) --- K-band flux in $10^{-14}$\,W~m$^{-2}\,\mu$m$^{-1}$, no error
             estimation is quoted for this measurement since the dominant
             source of uncertainty is that of the flux calibration.  The
             uncertainty in the flux calibration is $\sim$20\% for all of the
             sources.
Col.~(3) --- Peak K-band surface brightness in Figure~3, in units of
             10$^{-14}$\,W~m$^{-2}$~$\mu$m$^{-1}$~arcsec$^{-2}$.  The value quoted
             for NGC\,3393 corresponds to the KL-band measurement.
             For Mkn\,463W and Mkn\,463E
             only one value is given as they are both in the same field of view.  The
             two values cited for Circinus correspond to the K- and KL-band measurements,
             respectively.
Col.~(4) --- Peak Br$\gamma$ surface brightness in Figure~3, in units of
             10$^{-18}$\,W~m$^{-2}$~arcsec$^{-2}$.
Col.~(5) --- Equivalent width of Br$\gamma$ at 2.16\,$\mu$m in units of \AA\@.
             In case of no detection 3\,$\sigma$ upper limits are given.
Col.~(6) --- 1\,$\sigma$ estimated uncertainty in the equivalent width of
             Br$\gamma$ in \AA\@.
Col.~(7) --- Absolute flux of the line in units $10^{-18}$\,W~m$^{-2}$.  For
             Mrk\,1044, NGC\,4639 and NGC\,1194 (marked with asterisk), no
             calibration was possible.  The values shown in those cases are
             relative to the counts measured in the 2~arcsecond aperture spectra
             after those spectra were normalized to unity at 2.2\,$\mu$m.
             Values for these sources are shown in order to provide some
             measure of the change in flux with aperture.  In case of no
             detection 3\,$\sigma$ upper limits are given.
             For NGC\,4639 the position of the Br$\gamma$ line is at the blue limit
             of the spectrum so no values are given.
Col.~(8) --- 1\,$\sigma$ uncertainties estimated for the flux estimates given in
             Col.~(7).
Col.~(9) --- First order moment for the line in units of km~s$^{-1}$.  This
             provides a flux weighted estimate of the redshift without the
             assumption of a profile shape (such as a Gaussian).
Col.~(10) --- 1\,$\sigma$ uncertainty in the first moment.
Col.~(11) --- Estimate of the Gaussian FWHM, $\Gamma$, derived from the second
             order moment using the standard relation $\Gamma=2.355\sigma$, in
             units of km~s$^{-1}$.  The values have been corrected for
             instrumental broadening.  In cases of unresolved lines, no value
             is given.
Col.~(12) --- 1\,$\sigma$ uncertainty in the second moment.
$^\dagger$The absolute calibration is uncertain, since the our field-of-view is
smaller than the aperture listed in Column~10 of Table~5.

%Some of the object spectra deserve special note.  We do not
%report the redshifts for the AAT spectrum of NGC\,7582 and NGC\,3079 due to
%difficulties in those cases with the spectral calibration.  For NGC\,1097,
%NGC\,1433 and NGC\,1672 only K-band fluxes are reported.  The Br$\gamma$
%features seen in these spectra could possibly arise exclusively from residuum in
%substraction of Br$\gamma$ of the calibration star.  In these cases, it is impossible to
%determine whether these galaxies have significant Br$\gamma$ flux.  In
%the case of the western component of Mrk\,463, Br$\gamma$ is very weak and
%it is affected by the H-band second order contamination, in that
%observation since an order separating filter was not placed in front of
%the grism. Thus no value is reported for the western component of Mrk\,463.}

\begin{deluxetable}{cccccccccc}
\tablecolumns{9}
\tablewidth{0pt}
\tabletypesize{\scriptsize}
\tablenum{7}
\tablecaption{Properties of the K-band \mbox{H$_2$~1-0~S(1)} Emission Line}
\tablehead{
\colhead{Galaxy}&\colhead{Peak H$_2$}&\colhead{EW(H$_2$)}&
\colhead{$\sigma_{EW}$}&\colhead{Flux}&
\colhead{$\sigma_{Flux}$}&\colhead{cz}&
\colhead{$\sigma_{cz}$}&\colhead{W}&
\colhead{$\sigma_{W}$} \\
\colhead{(1)}&\colhead{(2)}&
\colhead{(3)}&\colhead{(4)}&
\colhead{(5)}&\colhead{(6)}&
\colhead{(7)}&\colhead{(8)}&
\colhead{(9)}&\colhead{(10)}}
\startdata
Mrk\,348     & \nodata & 3.3     & 0.5     & 0.56    & 0.07     & 4360    & 30      & 300     & 40     \\
 2           &         &  3.1    & 0.4     & 1.2     & 0.2      & 4340    & 30      & 340     & 60     \\
 4           &         & 3.2     & 0.6     & 2.0     & 0.4      & 4230    & 40      & 340     & 70 \\
I\,Zw\,1     & \nodata & $<$0.42 & \nodata & $<$0.83 & \nodata  & \nodata & \nodata & \nodata & \nodata \\
 2           &         & $<$0.46 & \nodata & $<$1.7  & \nodata  & \nodata & \nodata & \nodata & \nodata \\
 4           &         & $<$0.56 & \nodata & $<$2.5  & \nodata  & \nodata & \nodata & \nodata & \nodata \\
Mrk\,573     & \nodata & 2.7     & 0.4     & 3.1     & 0.5      & 5080    & 50      & 170     & 30     \\
 2           &         & 2.6     & 0.5     & 4.0     & 0.7      & 5080    & 90      & 120     & 20 \\
 4           &         & 2.3     & 0.5     & 7.9     & 1.7      & 5130    & 70      & 220     & 50     \\
Mrk\,1044*   & \nodata & $<$0.93 & \nodata & $<$0.62 & \nodata  & \nodata & \nodata & \nodata & \nodata \\
 2           &         & $<$1.2  & \nodata & $<$1.1  & \nodata  & \nodata & \nodata & \nodata & \nodata \\
 4           &         & $<$1.1  & \nodata & $<$1.7  & \nodata  & \nodata & \nodata & \nodata & \nodata \\
NGC\,1068$^\dagger$    &  9.9    & 0.66    & 0.09    & 5.1      & 0.7     & 1010    & 40      & 330     & 50    \\
 2           &         & 1.3     & 0.1     & 28      & 2.4      & 1052    & 20      & 310     & 30    \\
 4           &         & 2.6     & 0.1     & 75      & 4.0      & 1067    & 15      & 270     & 20     \\
NGC\,1097    & \nodata & 3.0     & 0.5     & 1.1     & 0.2      & 1160    & 30      & 470     & 80     \\
 2           &         & 2.6     & 0.4     & 2.8     & 0.4      & 1150    & 40      & 490     & 80     \\
 4           &         & 1.9     & 0.3     & 5.6     & 1.0      & 1110    & 50      & 510     & 90     \\
NGC\,1194*   & \nodata & $<$0.71 & \nodata & $<$0.77 & \nodata  & \nodata & \nodata & \nodata & \nodata \\
 2           &         & $<$1.1  & \nodata & $<$1.1  & \nodata  & \nodata & \nodata & \nodata & \nodata \\
 4           &         & $<$0.84 & \nodata & $<$2.0  & \nodata  & \nodata & \nodata & \nodata & \nodata \\
NGC\,1275    & 53      & 53      & 1.0     & 15.8    & 0.2      & 5090    & 5       & 410     & 11      \\
 2           &         & 53      & 0.8     & 25.2    & 0.3      & 5096    & 5       & 405     & 10     \\
 4           &         & 40      & 0.9     & 35.7    & 0.7      & 5103    & 5       & 394     &  12    \\
NGC\,1365$^\dagger$    & \nodata & $<$0.38 & \nodata & $<$0.58 & \nodata  & \nodata & \nodata & \nodata & \nodata \\
 2           &         & $<$0.44 & \nodata & $<$2.1  & \nodata  & \nodata & \nodata & \nodata & \nodata \\
 3           &         & $<$0.48 & \nodata & $<$3.0  & \nodata  & \nodata & \nodata & \nodata & \nodata \\
NGC\,1386    & \nodata &    2.2  & 0.4     & 1.4     & 0.2      & 780     & 35      & 280     & 40     \\
 2           &         &    2.0  & 0.3     & 2.0     & 0.3      & 760     & 25      & 260     & 40     \\
 4           &         &    1.3  & 0.2     & 3.4     & 0.6      & 680     & 25      & 240     & 40     \\
NGC\,1433$^\dagger$    & \nodata & $<$0.87 & \nodata & $<$0.14 & \nodata  & \nodata & \nodata & \nodata & \nodata \\
 2           &         & $<$0.87 & \nodata & $<$0.26 & \nodata  & \nodata & \nodata & \nodata & \nodata \\
 4           &         & $<$0.69 & \nodata & $<$0.56 & \nodata  & \nodata & \nodata & \nodata & \nodata \\
NGC\,1566    & \nodata & 1.8     & 0.2     & 0.65    & 0.09     & 1510    & 30      & 130     & 20     \\
 2           &         & 1.7     & 0.2     & 2.8     & 0.3      & 1504    & 14      & 190     & 20      \\
 4           &         & 1.8     & 0.2     & 6.5     & 0.7      & 1490    & 18      & 140     & 15    \\
NGC\,1672    & \nodata & $<$0.45 & \nodata & $<$0.18 & \nodata  & \nodata & \nodata & \nodata & \nodata \\
 2           &         & $<$0.35 & \nodata & $<$0.42 & \nodata  & \nodata & \nodata & \nodata & \nodata \\
 4           &         & $<$0.39 & \nodata & $<$1.2  & \nodata  & \nodata & \nodata & \nodata & \nodata \\
Mrk\,1095    & \nodata & $<$0.57 & \nodata & $<$0.63 & \nodata  & \nodata & \nodata & \nodata & \nodata \\
 2           &         & $<$0.56 & \nodata & $<$1.1  & \nodata  & \nodata & \nodata & \nodata & \nodata \\
 4           &         & $<$0.57 & \nodata & $<$1.6  & \nodata  & \nodata & \nodata & \nodata & \nodata \\
NGC\,2110    & \nodata & 3.4     & 0.3     & 0.85    & 0.08     & 2280    & 15      & 190     & 20     \\
 2           &         & 4.1     & 0.4     & 5.8     & 0.6      & 2270    &  20     & 280     & 30     \\
 4           &         & 6.1     & 0.7     & 15      & 1.7      & 2340    &  30     & 415     & 50     \\
NGC\,3079    & \nodata & 13      & 0.9     & 22      & 1.3      & 1009    & 25      & 447     & 30     \\
 2           &         & 13      & 1.6     & 20      & 2.3      & 1005    & 30      & 440     & 50     \\
 4           &         & 13      & 1.1     & 45      & 3.7      & 1008    & 25      & 420     & 30     \\
NGC\,3227    & 14      & 4.6     & 0.3     & 6.0     & 0.4      & 1170    & 20      & 253     & 20     \\
 2           &         & 4.8     & 0.2     & 14      & 0.7      & 1170    & 11      & 245     & 12     \\
 4           &         & 5.4     & 0.2     & 27      & 1.1      & 1170    & 15      & 245     & 10     \\
NGC\,3783    & \nodata & $<$0.53 & \nodata & $<$0.82 & \nodata  & \nodata & \nodata & \nodata & \nodata \\
 2           &         & $<$0.58 & \nodata & $<$1.2  & \nodata  & \nodata & \nodata & \nodata & \nodata \\
 4           &         & $<$0.66 & \nodata & $<$2.4  & \nodata  & \nodata & \nodata & \nodata & \nodata \\
NGC\,4051$^\dagger$    & \nodata & 1.9     & 0.2     & 1.8     & 0.2      & 632     & 30      & 250     & 30     \\
 2           &         & 2.0     & 0.2     & 4.2     & 0.5      & 612     & 25      & 210     & 25     \\
 4           &         & 2.2     & 0.2     & 6.7     & 0.7      & 625     & 20      & \nodata & \nodata  \\
NGC\,4258    & \nodata & $<$0.56 & \nodata & $<$0.42 & \nodata  & \nodata & \nodata & \nodata & \nodata \\
 2           &         & $<$0.59 & \nodata & $<$1.0  & \nodata  & \nodata & \nodata & \nodata & \nodata \\
 4           &         & $<$0.54 & \nodata & $<$2.2  & \nodata  & \nodata & \nodata & \nodata & \nodata \\
Mrk\,231     & 20      & 0.69    & 0.1     & 2.8     & 0.6      & 12660   & 30      & \nodata & \nodata \\
 2           &         & 0.80    & 0.1     & 6.1     & 1.0      & 12660   & 30      & 130     & 20      \\
 4           &         & 1.0     & 0.1     & 9.0     & 1.0      & 12650   & 25      & 140     & 15     \\
NGC\,4945    & 24      & 11.9    & 0.3     & 29      & 0.8      & 458     & 7       & 261     & 20     \\
 2           &         & 11.1    & 0.3     & 18      & 0.5      & 458     & 8       & 258     & 26     \\
 4           &         & 12.3    & 0.3     & 58      & 1.4      & 458     & 6       & 260     & 19      \\
Mrk\,273$^\dagger$     &  8.2    & 20      & 1.8     & 1.4     & 0.1      & 11180   & 30      & 600     & 90     \\
 2           &         & 19      & 1.2     & 4.2     & 0.2      & 11150   & 20      & 580     & 50     \\
 4           &         & 16      & 1.3     & 9.3     & 0.7      & 11120   & 50      & 510     & 60     \\
Circinus     & 19      & 2.7     & 0.2     & 4.3     & 0.3      & 445     & 10      & 265     & 20     \\
 2           &         & 3.6     & 0.2     & 23      & 1.3      & 450     & 8       & \nodata & \nodata \\
 3           &         & 3.5     & 0.2     & 35      & 2.4      & 448     & 10      & \nodata & \nodata \\
NGC\,5728$^\dagger$    & \nodata & 7.3     & 1.0     & 1.5     & 0.2      & 2860    & 50      & 380     & 50     \\
 2           &         & 7.3     & 1.1     & 2.6     & 0.4      & 2870    & 40      & 380     & 60     \\
 4           &         & 6.3     & 0.8     & 7.3     & 1.0      & 2890    & 40      & 390     & 50     \\
NGC\,7469    &  5.9    & 2.5     & 0.2     & 3.0     & 0.2      & 4750    & 13      & 325     & 20     \\
 2           &         & 2.6     & 0.1     & 6.8     & 0.4      & 4735    & 15      & 285     & 20     \\
 4           &         & 3.2     & 0.1     & 14      & 0.7      & 4715    & 10      & 260     & 15      \\
Mrk\,315     &  2.2    & $<$1.7  & \nodata & $<$0.16 & \nodata  & \nodata & \nodata & \nodata & \nodata \\
 2           &         & $<$1.6  & \nodata & $<$0.34 & \nodata  & \nodata & \nodata & \nodata & \nodata \\
 4           &         & 3.6     & 0.6     & 1.7     & 0.3      & 11500   & 40      & 430     & 70     \\
NGC\,7582(E) & \nodata & 1.5     & 0.2     & 2.6     & 0.4      & 1690    & 30      & 320     & 60     \\
 2           &         & 1.5     & 0.2     & 4.6     & 0.8      & 1690    & 40      & 310     & 70     \\
 4           &         & 2.0     & 0.2     & 14      & 1.6      & 1660    & 30      & 280     & 60     \\
NGC\,7582(A) &  3.8    & 1.4     & 0.2     & 2.9     & 0.5      & 1440    & 70      & 540     & 90     \\
 2           &         & 1.4     & 0.2     & 6.1     & 0.9      & 1490    & 40      & 410     & 60     \\
 4           &         & 2.2     & 0.2     & 16      & 1.5      & 1520    & 30      & 350     & 30
\tablecomments{
Col.~(1) --- Source designation.  The three rows per source have the same
             meaning as for Table~6.  Two sets of data are listed for NGC\,7582.
             The first are results obtained from the ESO 2.2m data and the
             second are results from the AAT data.
Col.~(2) --- Peak \mbox{H$_2$~1-0~S(1)} surface brightness in Figure~3, in units of
             10$^{-18}$\,W~m$^{-2}$~arcsec$^{-2}$.
Col.~(3) --- Equivalent width of \mbox{H$_2$~1-0~S(1)} in
             units of \AA\@.  In case of no detection 3\,$\sigma$ upper limits
             are given.
Col.~(4) --- 1\,$\sigma$ estimated uncertainty in the equivalent width of
             \mbox{H$_2$~1-0~S(1)} in \AA\@.
Col.~(5) --- Absolute flux of the line in units $10^{-18}$\,W~m$^{-2}$.  For
             Mrk\,1044 and NGC\,1194 (marked with asterisk), no
             calibration was possible.  The values shown are relative to the
             counts measured in the 2~arcsecond aperture spectra after those
             spectra were normalized to unity at 2.2\,$\mu$m.  These values are
             shown in order to provide some measure of the change in flux with
             aperture.  In case of no detection 3\,$\sigma$ upper limits are
             given.
Col.~(6) --- Uncertainties estimated for the flux estimates given in Col.~(5).
Col.~(7) --- First order moment for the line in units of km~s$^{-1}$.
Col.~(8) --- 1\,$\sigma$ uncertainty in the first moment.
Col.~(9) --- Estimate of the Gaussian FWHM, $\Gamma$, derived from the second
             order moment using the standard relation $\Gamma=2.355\sigma$, in
             units of km~s$^{-1}$.  The values have been corrected for
             instrumental broadening.  In cases of unresolved lines, no value
             is given.
Col.~(10) --- 1\,$\sigma$ uncertainty in the second moment.
$^\dagger$The absolute calibration is uncertain, since the our field-of-view is
smaller than the aperture listed in Column~10 of Table~6.
%In the case of the western component of Mrk\,463, the
%spectrum is affected by the H-band second order contamination, in that
%observation since an order separating filter was not placed in front of
%the grism.  Thus no value is reported for the western component of Mrk\,463.
}
\enddata
\end{deluxetable}

\begin{deluxetable}{cccccccccccc}
\tablecolumns{10}
\tablewidth{0pt}
\tabletypesize{\scriptsize}
\tablenum{8}
\tablecaption{H-band Flux and properties of the [Fe{\tiny II}]1.644$\mu$m Emission Line}
\tablehead{
\colhead{Galaxy}& \colhead{H-band Flux}&
\colhead{H-band Peak}&\colhead{[Fe{\tiny II}] Peak}&
\colhead{EW([Fe{\tiny II}])}&\colhead{$\sigma_{EW}$}&
\colhead{Flux}&\colhead{$\sigma_{Flux}$}&
\colhead{cz}&\colhead{$\sigma_{cz}$}&
\colhead{W}& \colhead{$\sigma_{W}$} \\
\colhead{(1)}&\colhead{(2)}&
\colhead{(3)}&\colhead{(4)}&
\colhead{(5)}&\colhead{(6)}&
\colhead{(7)}&\colhead{(8)}&
\colhead{(9)}&\colhead{(10)}&
\colhead{(11)}&\colhead{(12)}}
\startdata
I\,Zw\,1     & 1.8  &  2.0     & \nodata & $<$0.56  & \nodata & $<$1.1   & \nodata & \nodata &  \nodata & \nodata & \nodata \\
 2           & 3.1  &          &         & $<$0.57  & \nodata & $<$1.9   & \nodata & \nodata &  \nodata & \nodata & \nodata \\
 4           & 4.4  &          &         & $<$0.57  & \nodata & $<$2.7   & \nodata & \nodata &  \nodata & \nodata & \nodata \\
Mrk\,1044*   & 0.76 & \nodata  & \nodata & $<$0.37  & \nodata & $<$0.28  & \nodata & \nodata &  \nodata & \nodata & \nodata \\
 2           & 1.0  &          &         & $<$0.38  & \nodata & $<$0.40  & \nodata & \nodata &  \nodata & \nodata & \nodata \\
 4           & 1.9  &          &         & $<$0.39  & \nodata & $<$0.76  & \nodata & \nodata &  \nodata & \nodata & \nodata \\
NGC\,1068    & 7.1  & 10       & 47      & 3.7      & 0.1     & 28       & 1.1     & 1089    &  20      & 625     & 25 \\
 2           & 16   &          &         & 4.6      & 0.1     & 76       & 1.7     & 1100    &  10      & 663     & 20 \\
 4           & 30   &          &         & 6.9      & 0.1     & 211      & 3.9     & 1128    &  10      & 723     & 20 \\
NGC\,1097    & 0.67 &  0.87    & \nodata & $<$0.43  & \nodata & $<$0.29  & \nodata & \nodata &  \nodata & \nodata & \nodata \\
 2           & 2.0  &          &         & $<$0.36  & \nodata & $<$0.72  & \nodata & \nodata &  \nodata & \nodata & \nodata \\
 4           & 5.8  &          &         & $<$0.48  & \nodata & $<$2.7   & \nodata & \nodata &  \nodata & \nodata & \nodata \\
NGC\,1275    & 0.33 &  0.90    & 20      & 55       & 1.9     & 18.8     & 0.6     & 5042    &  18      & 760     & 30  \\
 2           & 0.53 &          &         & 54       & 1.8     & 29.5     & 0.8     & 5040    &  20      & 755     & 30   \\
 3           & 0.91 &          &         & 42       & 2.0     & 39       & 1.7     & 5030    &  20      & 765     & 30    \\
NGC\,1365$^\dagger$    & 2.4  &  6.8     & \nodata & $<$0.16  & \nodata & $<$0.39  & \nodata & \nodata &  \nodata & \nodata & \nodata \\
 2           & 6.0  &          &         & $<$0.20  & \nodata & $<$1.2   & \nodata & \nodata &  \nodata & \nodata & \nodata \\
 4           & 8.8  &          &         & $<$0.22  & \nodata & $<$1.9   & \nodata & \nodata &  \nodata & \nodata & \nodata \\
NGC\,1386    & 0.60 &  0.93    &  7.8    & 6.1      & 0.2     & 3.6      & 0.1     & 880     &  10      & 470     & 25 \\
 2           & 2.2  &          &         & 4.6      & 0.1     & 9.9      & 0.3     & 880     &   9      & 460     & 20 \\
 4           & 5.3  &          &         & 2.2      & 0.1     & 11       & 0.5     & 870     &  12      & 430     & 30  \\
NGC\,1433$^\dagger$    & 0.40 &  0.31    & \nodata & $<$0.44  & \nodata & $<$0.18  & \nodata & \nodata &  \nodata & \nodata & \nodata \\
 2           & 0.76 &          &         & $<$0.38  & \nodata & $<$0.29  & \nodata & \nodata &  \nodata & \nodata & \nodata \\
 4           & 2.1  &          &         & $<$0.47  & \nodata & $<$0.97  & \nodata & \nodata &  \nodata & \nodata & \nodata \\
NGC\,1566    & 0.80 &  1.5     &  5.0    & 2.5      & 0.1     & 2.0      & 0.1     & 1460    &  11      & 205     & 10 \\
 2           & 2.6  &          &         & 1.4      & 0.1     & 3.6      & 0.3     & 1444    &  14      & 148     & 10 \\
 4           & 5.7  &          &         & $<$0.77  & \nodata & $<$4.3   & \nodata & \nodata &  \nodata & \nodata & \nodata \\
NGC\,1672    & 0.24 &  0.94    & \nodata & $<$0.50  & \nodata & $<$0.12  & \nodata & \nodata &  \nodata & \nodata & \nodata \\
 2           & 1.9  &          &         & $<$0.42  & \nodata & $<$0.78  & \nodata & \nodata &  \nodata & \nodata & \nodata \\
 4           & 4.8  &          &         & $<$0.49  & \nodata & $<$2.3   & \nodata & \nodata &  \nodata & \nodata & \nodata \\
Mrk\,1095    & 1.4  &  6.0     & \nodata & $<$0.44  & \nodata & $<$0.62  & \nodata & \nodata &  \nodata & \nodata & \nodata \\
2            & 2.4  &          &         & $<$0.47  & \nodata & $<$1.1   & \nodata & \nodata &  \nodata & \nodata & \nodata \\
4            & 3.6  &          &         & $<$0.54  & \nodata & $<$1.9   & \nodata & \nodata &  \nodata & \nodata & \nodata \\
NGC\,2110    & 3.1  &  6.6     & 32      & 10.9     & 0.3     & 32.6     & 1.0     & 2220    &   7      & 406     & 13 \\
 2           & 1.9  &          &         & 14.3     & 0.4     & 26.2     & 0.7     & 2240    &   8      & 456     & 12 \\
 4           & 5.2  &          &         & 10.0     & 0.3     & 50.7     & 1.5     & 2220    &  10      & 388     & 11 \\
NGC\,3227    & 3.2  &  2.4     & 17      & 7.3      & 0.4     & 23       & 1.1     & 927     &  22      & 574     & 30 \\
 2           & 4.5  &          &         & 7.3      & 0.4     & 32       & 1.7     & 920     &  35      & 540     & 30 \\
 4           & 8.0  &          &         & 6.7      & 0.4     & 53       & 2.8     & 930     &  20      & 580     & 30 \\
NGC\,3783    & 1.0  &  0.82    & \nodata & 1.9      & 0.1     & 1.9      & 0.1     & 2840    &  15      & 137     & 40 \\
 2           & 1.7  &          &         & 1.8      & 0.2     & 3.0      & 0.3     & 2840    &  30      & 117     & 50 \\
 4           & 3.5  &          &         & 1.4      & 0.2     & 4.9      & 0.6     & 2840    &  20      & 190     & 25  \\
NGC\,4051$^\dagger$    & 1.5  &  3.3     & \nodata & 0.76     & 0.2     & 1.2      & 0.2     & 370     &  30      & 150     & 50   \\
 2           & 2.9  &          &         & 0.72     & 0.2     & 2.0      & 0.5     & 380     &  30      & 120     & 60   \\
 4           & 4.2  &          &         & $<$0.54  & \nodata & $<$2.2   & \nodata & \nodata &  \nodata & \nodata & \nodata \\
NGC\,4945    & 1.3  &  0.61    &  4.5    & 6.2      & 0.2     & 8.3      & 0.3     & 426     &  9      & 330     & 22 \\
 2           & 1.5  &          &         & 6.0      & 0.3     & 9.5      & 0.5     & 429     &  10      & 334     & 29 \\
 4           & 4.4  &          &         & 6.4      & 0.2     & 29.3     & 1.0     & 427     &  10      & 313     & 30 \\
NGC\,7469    & 2.3  &  5.6     &  9.7    & 1.7      & 0.2     & 3.7      & 0.4     & 4760    &  25      & 350     & 35  \\
 2           & 4.3  &          &         & 2.1      & 0.2     & 8.8      & 0.9     & 4750    &  15      & 290     & 30 \\
 4           & 6.8  &          &         & 2.9      & 0.2     & 19       & 1.1     & 4754    &  11      & 290     & 16 \\
NGC\,7582(A) & 2.0  &  7.8     &  5.1    & 2.5      & 0.1     & 5.1      & 0.2     & 1617    &  12      & 265     & 10 \\
 2           & 4.2  &          &         & 2.7      & 0.1     & 11.5     & 0.4     & 1614    &  10      & 288     & 10 \\
 4           & 9.4  &          &         & 3.5      & 0.1     & 33.6     & 1.1     & 1603    &  10      & 336     & 10

\tablecomments{
Col.~(1) --- Source designation.  The three rows per source have the same meaning as for Table~6.
Col.~(2) --- H-band flux in $10^{-14}$\,W~m$^{-2}$~$\mu$m$^{-1}$.  No error estimation is quoted
             for this measurement since the dominant sources of uncertainty is that of the flux
             calibration.  The uncertainty in the flux calibration is $\sim$20\% for all of the sources.
Col.~(3) --- Peak H-band surface brightness in Figure~3, in units of
             10$^{-14}$\,W~m$^{-2}$~$\mu$m$^{-1}$~arcsec$^{-2}$.  The values quoted
             for NGC\,4945, NGC\,7469, and NGC\,7582 correspond to the HH-band measurement.
Col.~(4) --- Peak [Fe{\tiny II}]$\lambda$1.644$\mu$m surface brightness in Figure~3, in units of
             10$^{-18}$\,W~m$^{-2}$~arcsec$^{-2}$.
Col.~(5) --- Equivalent width of the [Fe{\tiny II}]$\lambda$1.644$\mu$m emission line in units of \AA\@.
             In case of no detection 3\,$\sigma$ upper limits are given.
Col.~(6) --- 1\,$\sigma$ estimated uncertainty in the equivalent width of [Fe{\tiny II}]$\lambda$1.644$\mu$m in \AA\@.
Col.~(7) --- Absolute flux of the line in units $10^{-18}$\,W~m$^{-2}$.  For Mrk\,1044 no calibration was possible.
             The values shown are relative to the counts measured in the 2 arcsecond aperture spectrum after it was
             normalized to unity at 2.2\,$\mu$m and the continuum was then extrapolated to the H-band.  These values
             are shown in order to provide some measure of the change in flux with aperture and for allowing the
             comparison of features in the H- and K-band in Mrk\,1044.  In case of no detection 3\,$\sigma$ upper
             limits are given.
Col.~(8) --- Uncertainties estimated for the fluxes given in Col.~(7).
Col.~(9) --- First order moment for the line in units of km~s$^{-1}$.  This provides a flux weighted estimate of the
             redshift without the assumption of a profile shape (such as a Gaussian).  Some of the line profiles
             were apparently contaminated by absorption from the adjacent CO bandhead.  Where this occurs it did
             not appear to significantly change the flux, EQW, or the redshift estimates for the strongest lines.
Col.~(10) --- 1\,$\sigma$ uncertainty in the first moment.
Col.~(11) --- Estimate of the second order moment in units of km~s$^{-1}$.  The values have been corrected for
              instrumental broadening.
Col.~(12) --- 1\,$\sigma$ uncertainty in the second moment.
$^\dagger$The absolute calibration is uncertain, since the our field-of-view is
smaller than the aperture listed in Column~10 of Table~5.}

\enddata
\end{deluxetable}

\end{document}